\documentclass[aps,pre,twocolumn,groupedaddress,showpacs,superscriptaddress,amssymb,amsmath]{revtex4-1}
\usepackage{graphicx} 
\usepackage{hyperref}
\hypersetup{breaklinks,colorlinks,linkcolor=blue,citecolor=blue,urlcolor=blue}
\usepackage{color}
\usepackage{subfigure}
\newcommand{\be}{\begin{equation}}
\newcommand{\ee}{\end{equation}}
\newcommand{\bea}{\begin{eqnarray}}
\newcommand{\eea}{\end{eqnarray}}

\begin{document}

\title{Machine learning delta-T noise for temperature bias estimation}

\author{Matthew Gerry*}
\altaffiliation{These authors contributed equally to this work.}
\affiliation{Department of Physics, University of Toronto, 60 Saint George St., Toronto, Ontario M5S 1A7, Canada}

\author{Jonathan J. Wang*}
\altaffiliation{These authors contributed equally to this work.}
\affiliation{Department of Chemistry,
University of Toronto, 80 Saint George St., Toronto, Ontario, M5S 3H6, Canada}

\author{Joanna Li}
\affiliation{Department of Physics, University of Toronto, 60 Saint George St., Toronto, Ontario M5S 1A7, Canada}
\affiliation{Division of Engineering Science, University of Toronto, 42 Saint George St., Toronto, Ontario M5S 2E4, Canada}

\author{Ofir Shein-Lumbroso}
\affiliation{Department of Chemical and Biological Physics, Weizmann Institute of Science, Rehovot 7610001, Israel}

\author{Oren Tal}
\affiliation{Department of Chemical and Biological Physics, Weizmann Institute of Science, Rehovot 7610001, Israel}

\author{Dvira Segal}
\email{dvira.segal@utoronto.ca}
\affiliation{Department of Chemistry, 
University of Toronto, 80 Saint George St., Toronto, Ontario, M5S 3H6, Canada}
\affiliation{Department of Physics, University of Toronto, 60 Saint George St., Toronto, Ontario M5S 1A7, Canada}

\date{\today}

\begin{abstract} 
Delta-T shot noise is activated in temperature-biased electronic junctions, down to the atomic scale. It is characterized by a quadratic dependence on the temperature difference and a nonlinear relationship with the transmission coefficients of partially opened conduction channels.
In this work, we demonstrate that delta-T noise, measured across an ensemble of atomic-scale junctions, can be utilized to estimate the temperature bias in these systems.
Our approach employs a supervised machine learning algorithm to train a neural network with input features being the scaled electrical conductance, the delta-T noise, and the mean temperature. 
Due to limited experimental data, we generate synthetic datasets, designed to mimic experiments. The neural network, trained on these synthetic data, was subsequently applied to predict temperature biases from experimental datasets.
Using performance metrics, we demonstrate that the mean bias---the deviation of predicted temperature differences from their true value---is less than 1 K for junctions with conductance up to 4$G_0$.
Our study highlights that, while a single delta-T noise measurement is insufficient for accurately estimating the applied temperature bias due to noise contributions from other sources, averaging over an ensemble of junctions enables predictions within experimental uncertainties. This demonstrates that machine learning approaches can be utilized for estimation of temperature biases, and similarly other stimuli in electronic junctions.
\end{abstract}

\maketitle

\section{Introduction}


Machine learning (ML) methods have found diverse applications across various fields, including materials science, quantum science, and nanoscience, offering an array of tools such as supervised, unsupervised, and reinforcement learning, generative models, and quantum machine learning tools. In the field of Physical Chemistry, recent efforts have focused on leveraging ML for generating machine-learning potentials, accelerating electronic structure calculations, unraveling structure-function relations, predicting crystal structures, and more \cite{MLPC1,MLPC2}. 

Measurements of electron transport in single-molecule junctions can uncover information on the structure and dynamics of the conducting molecule \cite{RevMO}.
For example, single-molecule junctions can be used to probe electron-vibrational couplings and thermal relaxation rates \cite{Natelsonev,Orenev}, the characteristics of metal-molecule bonds \cite{LathaMM}, and the coupling of electron motion with the chiral structure \cite{OrenChiral}. Furthermore, monitoring the conductance of single molecules that undergo a chemical reaction can be used to assess important reactions and their mechanisms, such as catalytic processes \cite{LathaRX}.  

Over the past decade, a variety of machine learning methods have been applied to analyze and predict single-molecule conduction \cite{Gemma22,Taniguchi23}. 
These efforts can be broadly categorized into two types. 
First, since molecular transport experiments are typically conducted repeatedly over thousands of junctions and exhibit significant variability, many studies focus on assisting experimentalists with data filtering and classification. This includes the classification of conductance traces in break-junction experiments to automate the identification of different structures and group the traces into families \cite{Gemma18,Perrin19, Vladyka,Hong20,Hong21,Zant19,Zant21,Latha20,Gemma24}.
These studies rely on a variety of ML algorithms, such as supervised and unsupervised learning, image recognition, and clustering. 
A complementary effort in ML-based quantum transport studies addresses structure-function relationships, that is, the prediction of electrical conductance in molecular junctions using ML algorithms. Examples include predicting transport in nanostructures with scattering impurities \cite{Anatole14,Li20} and in double-stranded DNA with varying sequences \cite{RomanDNA,MaitiDNA}. Recent studies have further advanced ML workflows by incorporating first-principles calculations and molecular dynamic simulations to predict the electrical conductance of, for example, metal nanowires \cite{FP21} and organic molecules \cite{MDML,Franco23}.
More broadly, we also mention an emerging application of ML to quantum thermal machines, which is to optimize performance \cite{Ilia1,Erdman1,Erdman2,Erdman3}, and test the regime of validity of basic bounds that are non-universal \cite{Ilia2}.

\begin{figure}
    \centering
    \includegraphics[width=\linewidth]{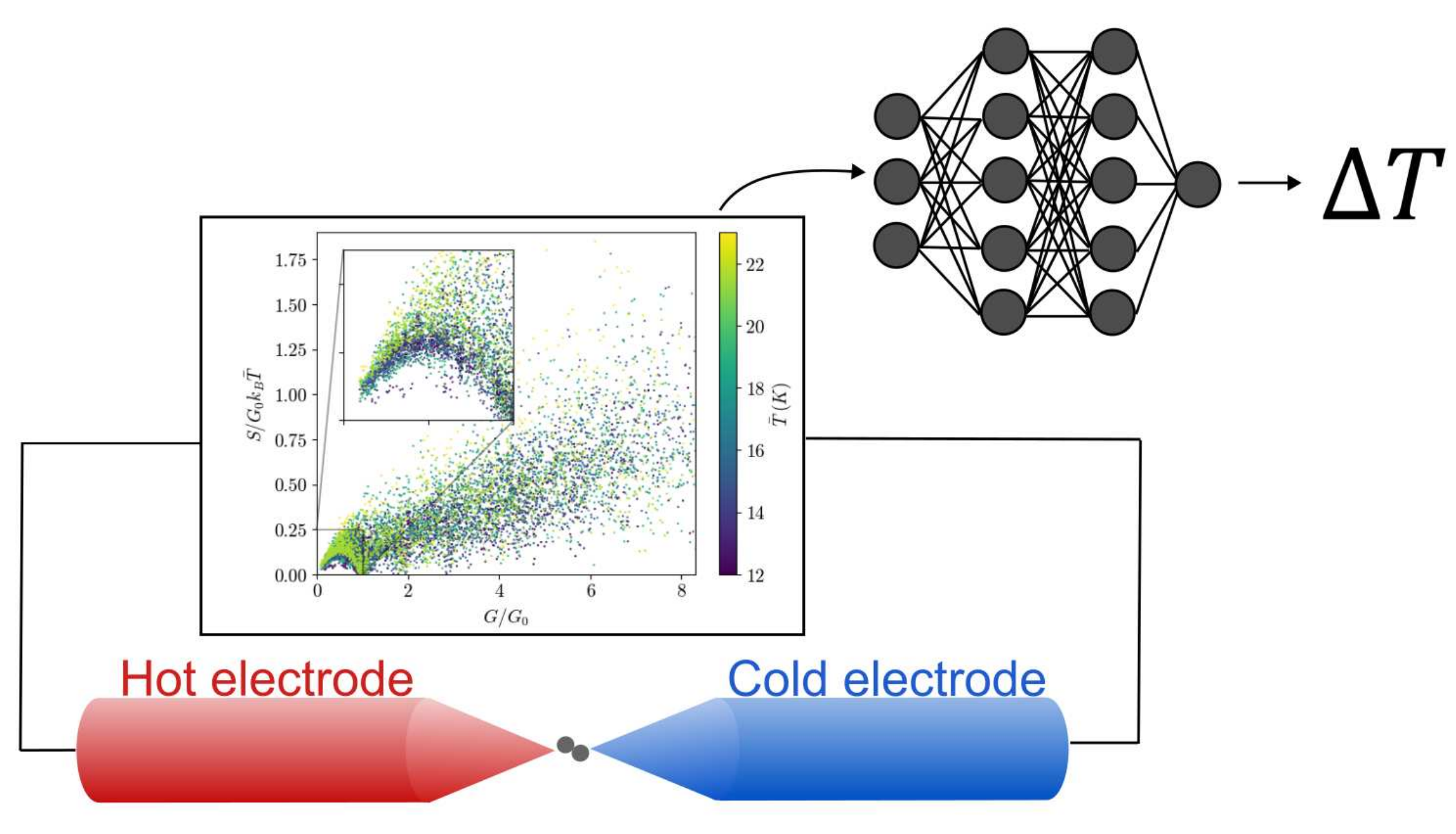}
    \caption{A schematic diagram of a single-molecule atomic-scale junction with a scatter plot showing the delta-T noise, $S_{\Delta T}$ against the conductance $G$. This diagram depicts the experiment of Ref. \citenum{DeltaTShot}.
    A neural network is employed here to estimate $\Delta T$ using temperature, conductance, and noise data.}
    \label{fig:setup_diagram}
\end{figure}

In this paper, we explore a novel application of ML methods to quantum transport. Specifically, we show that measurements of the electronic current noise in ensembles of atomic-scale junctions, combined with their electrical conductance, can be used via a supervised learning approach to infer the stimuli affecting transport. In the present case, we demonstrate that shot noise measurements enable estimation of the applied temperature bias \cite{DeltaTShot}. Our approach is represented in Fig. \ref{fig:setup_diagram} along with a schematic diagram of a single-molecule junction with a temperature bias.

As part of this study, we generate synthetic datasets of electrical conductance and the corresponding delta-T noise to supplement experimental data. Models trained on these synthetic datasets are subsequently tested on experimental results.
The process of reducing prediction errors on real datasets provides valuable insights into the sequential channel-opening protocol, enabling the development of a model that reasonably aligns with the observed behavior in real junctions.



The paper is organized as follows.
In Sec. \ref{sec:DeltaT}, we review the characteristics of the delta-T noise, elaborate on our research objective, and describe available experimental data collected in atomic-scale junctions. 
Given the limitation of the experimental datasets, in Sec. \ref{sec:Syn} we describe protocols for generating synthetic data. We present two approaches for conductance channel opening: a deterministic protocol, which proves inconsistent with experimental results, and a noisy protocol, which successfully captures measurements.
In Sec. \ref{sec:ML}, we describe the ML model developed for temperature bias estimation, including the training and testing procedures. We summarize our findings in Sec. \ref{sec:Summ} and offer future directions. Details of channel opening protocols are described in Appendix \ref{sec:App0}.
The architecture of the neural network (NN) is described in Appendix \ref{sec:App1}. Training and testing of the model using the delta-T integral noise formula are presented in Appendix \ref{sec:App2}.

\section{Delta-T shot noise}
\label{sec:DeltaT}

\subsection{Noise characteristics}
When a temperature difference $\Delta T$ is applied across a quantum conductor, current shot noise is generated. This delta-T noise was first demonstrated in an atomic-scale junctions \cite{DeltaTShot} and later observed in mesoscopic conductors \cite{TunnelJ}. 
Theoretically, using the full counting statistics approach of coherent transport \cite{Levitov1,Levitov2,Rev1},
it was shown in Ref. \citenum{DeltaTShot} that the delta-T noise can be approximated by 
\bea
\label{eq:S}
    S_I &=&4G_0k_B\bar T \sum_i\tau_i   
    \nonumber\\
    &+& G_0k_B\frac{\Delta T^2}{\bar{T}}\left(\frac{\pi^2}{9}-\frac{2}{3}\right) \sum_i\tau_i(1-\tau_i).
\eea
Here, $G_0=2e^2/h$ is the quantum of conductance, $\bar T$ is the average of $T_h$ and $T_c$, which are the temperatures at the hot and cold terminals of the conductor, and $\Delta T=T_h-T_c$. The noise depends on the transmission constants $\tau_i$ of the $i$th transmission channel. It is useful to recall that the electrical conductance itself is given in terms of the transmission values as
\begin{equation}
G=G_0\sum_i \tau_i.
   \end{equation}
Appendix \ref{sec:App2} provides a summary of the derivation of Eq. (\ref{eq:S}).
In short, the theory assumes noninteracting coherent electron transport. Additional steps leading to Eq. (\ref{eq:S}) are
(i) assuming constant (energy independent) transmission functions, and
(ii) expanding the noise to the lowest nontrivial $\Delta T/\bar T$ contribution, which is quadratic. 
As shown in Ref. \citenum{DeltaTShot},  
The second order expression proved to be a good approximation for the noise even when $\Delta T/\bar T$ was of the order of 1 \cite{DeltaTShot}.

The first term in Eq. (\ref{eq:S}) is analogous in form to the Johnson-Nyquist equilibrium noise, here evaluated at the {\it average} temperature across the conductor.  The second contribution, termed ``excess noise", and specifically ``delta-T  noise", is denoted by $S_{\Delta T}$, and is generated as a result of the temperature difference across the conductor. This contribution depends on the distribution of the transmission function across $N$ channels as $\sum_{i=1}^N \tau_i(1-\tau_i)$. That is, delta-T noise is generated by having {\it partial} transmission and backscattering of electrons.

Following experiments \cite{DeltaTShot,TunnelJ}, delta-T shot noise has been theoretically analyzed in different regimes, extending beyond the quantum coherent and constant transmission model. These studies examined the static \cite{AnqiR,qdot,Janine1} and finite-frequency \cite{light,Janine2} resonant transport limit, hybrid normal-superconducting systems \cite{SuperC}, as well as correlated electron junctions \cite{Hall, Kondo,FHall,color}.
Interestingly, the delta-T noise, a nonequilibrium noise induced by a temperature difference, was generated in the setup of Ref. \citenum{DeltaTShot} in the absence of net charge current due to the cancellation of equal and opposite currents flowing above and below the chemical potential. This novel type of zero-current nonequilibrium noise was recently proved to be bounded by equilibrium noise \cite{Janine1,Janine2}.


Numerous studies have shown that {\it voltage-activated} shot noise can probe the structural and electronic properties of atomic-scale and molecular junctions \cite{Rev2,Exp}. For example, noise measurements have been employed to resolve conduction channels \cite{Vardimon2013, Berndt19,Berndt20,Sheer16,Sheer21}, identify and probe interactions \cite{Natelson1,Natelson2} possibly due to effects at the interface \cite{Latha15,Anqi} or intrinsic to the conductor, such as electron-vibration interactions \cite{Nitzanev,lightN,Bijayev,Orenev}, and
characterize transmission properties \cite{Latha24}.

In this study, we continue these efforts to use noise as a probe for nanoscale transport phenomena. Beyond the fundamental significance of observing and characterizing delta-T noise, we demonstrate that this contribution, when combined with knowledge of thermal noise, can serve as a probe for detecting temperature differences in nanoscale and atomic-scale quantum conductors. 
However, to utilize the delta-T noise for this purpose, one needs to know the specific combination of transmission coefficients that construct this partition-type noise, as described in Eq. (\ref{eq:S}). This dependency renders solving the inverse problem particularly challenging.
To address this problem, we employ a supervised machine learning approach.
By inputting $G$, $S_{\Delta T}$, and $\bar T$, we show that a neural network can be trained to predict the temperature difference, $\Delta T$.
An important finding is that temperature difference prediction is valuable only when repeated across an ensemble of nanojunctions. 

In Sec. \ref{sec:Exp}, we describe the experimental characterization of delta-T noise in atomic-scale junctions \cite{DeltaTShot}. However, to convincingly demonstrate the ML prediction of temperature biases based on delta-T noise, we supplement experimental datasets with synthetic data, a task that we follow in Sec. \ref{sec:Syn}.


\begin{figure*}[htbp]
    \centering
    \includegraphics[width=1.0\linewidth]{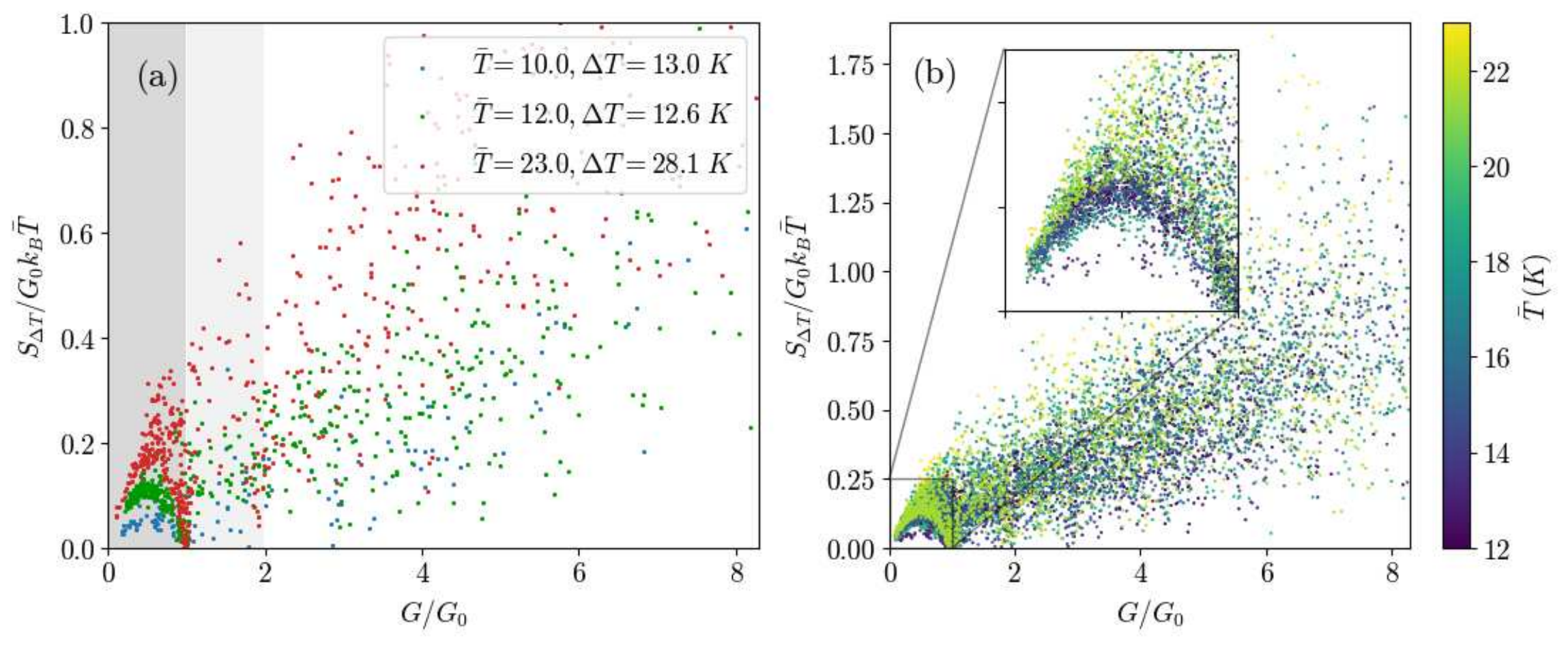}
    \caption{Visual representation of experimental measurements. (a) Scaled shot noise $S_{\Delta T}$ is plotted against electrical conductance $G$ in units of the conductance quantum $G_0$ for three choices of ($\bar{T}, \Delta T$) pairs used in the experiment. Two regions of interest, $0<G<G_0$ and $G_0<G<2G_0$, are shaded. In the former, quadratic dependence of $S_{\Delta T}$ on $G$ is observed with some noise. In the latter, this behavior becomes less pronounced. 
    Overall, $S_{\Delta T}$ tends to be higher at higher $\Delta T/\bar{T}$. (b) All the collected experimental data (excluding points taken at very low $\Delta T$) with the color bar indicating the average temperature $\bar{T}$ of the measurement environment. For this dataset, the temperature difference $\Delta T$ between the sides of the junction is highly similar in value to its corresponding $\bar{T}$, thus the colors can represent $\Delta T$ as well. The data presented in (a) is a subset of datasets in (b), and we display it in two different ways to highlight its dependence on $\Delta T$ and $\bar T$.}
    \label{fig:exp}
\end{figure*}
\begin{figure}[htpb]
    \centering
    \includegraphics[width=1\linewidth]{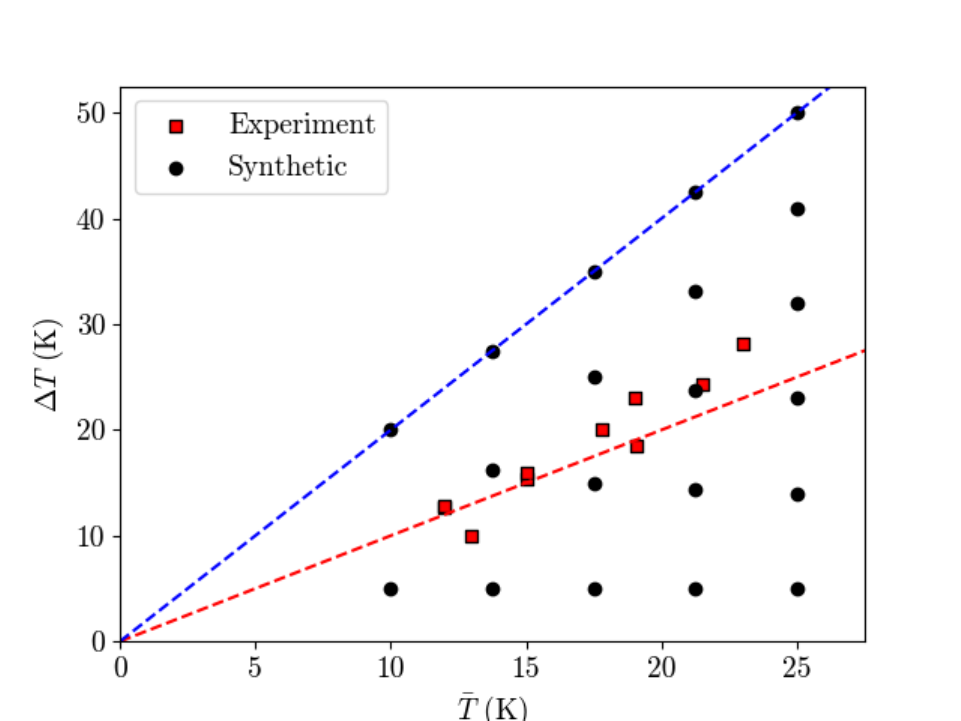}
    \caption{Pairs of $\Delta T$ and $\bar T$ of experimental datasets (red squares), and compared to $\Delta T$ and $\bar{T}$ values chosen to generate synthetic data (black circles). The red dashed line represents $\Delta T = \bar{T}$, in proximity to which the experimental data lies. Synthetic $\Delta T$ values range to the maximally allowed $\Delta T = 2\bar{T}$, with the border indicated by the dashed blue line.}
    \label{fig:deltaT-T}
\end{figure}

\begin{figure}[htbp]
    \centering
    \includegraphics[width=\linewidth]{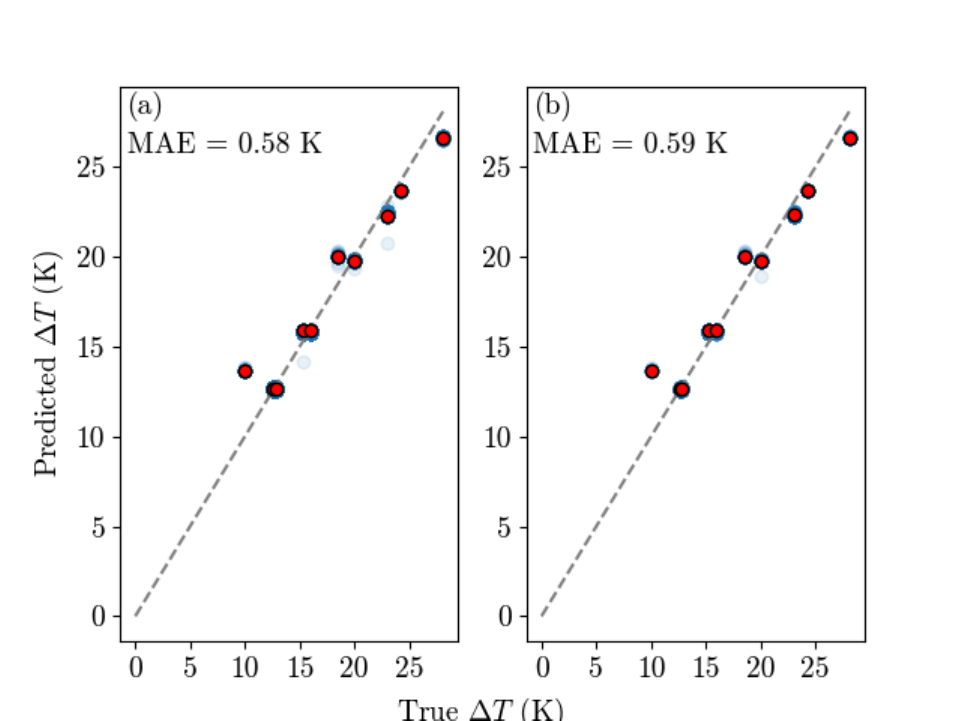}
    \caption{Neural network predicted $\Delta T$ plotted against true $\Delta T$ for (a) training set and (b) testing set from a model trained with experimental data. The dataset is split as 80\% training set and 20\% testing set. The red circles indicate the mean of all predicted $\Delta T$ (in blue) for an unique true $\Delta T$ with the standard deviation as error bars.
    }
    \label{fig:NNE}
\end{figure}

\subsection{Experimental data of delta-T noise and its limitations}
\label{sec:Exp}

The delta-T shot noise was experimentally characterized in Ref. \citenum{DeltaTShot} using atomic-scale junctions, 
where hydrogen molecules were introduced into the contact between two atomically-sharp gold electrodes.
Unlike bare gold atomic junctions, hydrogen-based molecular junctions offer a wide range of conductance values below 1$G_0$. This was crucial because the primary effort in Ref. \citenum{DeltaTShot} has been to characterize the delta-T shot noise in the conductance range of $G\leq 1 G_0$.

In brief, the experiment reported in Ref. \citenum{DeltaTShot} utilized the break junction technique to form an ensemble of molecular junctions with different structures at the contact region, resulting in a broad range of conductance values. A temperature gradient was applied on the junction by an asymmetric heating of the electrodes. The temperature bias across the junction was monitored by thermometers placed on opposite sides of the junction. Details over sample fabrication, the break junction technique, electronic measurements, junctions' characterization, calibration of thermometers, and delta-T noise measurements were included in Ref. \citenum{DeltaTShot}.

In the experiment, the current shot noise $S_I$ was measured. The delta-T noise $S_{\Delta T}$, an excess noise, is defined as the total noise minus the average thermal noise, which is determined by the average temperature, leading to the expression $S_{\Delta T} = S_I-4G k_B\bar T$.

In Fig. \ref{fig:exp}, we present examples of experimental datasets of delta-T noise as a function of the junction conductance. 
Each marker presents the noise measured on a single atomic-scale junction, with these measurements repeated across an ensemble of junctions. The data was collected under different average junction temperatures and for different temperature differences. 
The dark gray region highlights data up to 1$G_0$. 
It is important to note that only delta-T noise within this range was analyzed in Ref. \citenum{DeltaTShot}. This distinction is important because our present study extends the analysis beyond this conductance regime.
While the $G<1G_0$ data manifest clear signatures of partition noise characteristics, $S_{\Delta T}\propto \tau(1-\tau)$, given the dominance of a single channel in this regime, this trend is difficult to discern beyond 1$G_0$. We highlight in light gray the region of 
$1G_0<G<2G_0$. The partition noise characteristics are difficult to discern already at this range. 
Consequently, the verification of Eq. (\ref{eq:S}) for the full range of conductance has not been assessed so far.

The collected experimental data as presented in Fig. \ref{fig:exp}(b) includes 9 sets of ($\bar T$, $\Delta T$) pairs, for which both conductance $G$ and shot noise $S_I$ were measured for ensemble of junctions, the latter allowing the extraction of $S_{\Delta T}$.
In Fig. \ref{fig:deltaT-T}, we mark these 9 pairs by red squares. 

The results of training and testing an ML model on the experimental data are presented in Figure \ref{fig:NNE}; Details of training and testing are provided in Sec. \ref{sec:ML}.
The NN demonstrates an excellent ability to predict the temperature bias on the basis of the measured data. Although this is certainly encouraging, it is evident that, due to constraints in the experimental setup, there is a clear linear correlation between the variables $\bar T$ and $\Delta T$, as shown in Fig. \ref{fig:deltaT-T}. 
As a result of this correlation, the NN effectively learned that $\Delta T$ was restricted around $\bar T$, enabling accurate predictions of temperature biases without the need to deeply learn the characteristics of the noise.
To more carefully and broadly test the ability of an NN model to learn temperature differences from delta-T shot noise in molecular junctions, in the next section, we turn to synthetic data.


\begin{figure}[htbp]
    \centering
    \includegraphics[width=1.1\linewidth]{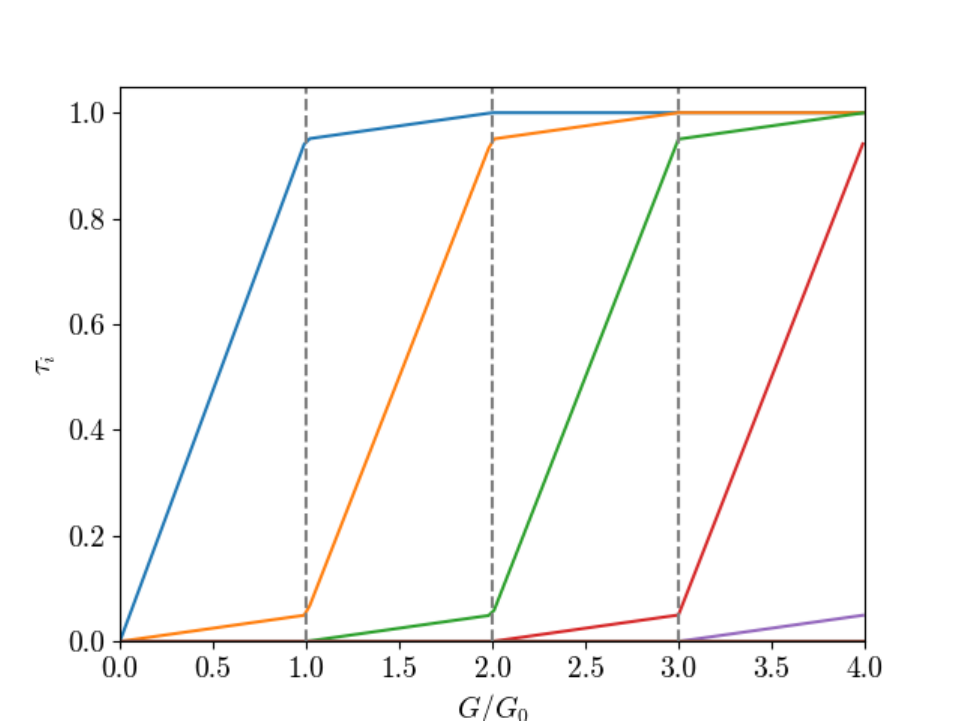}
    \caption{The transmission probability $\tau_i$ when successively opening channels within the deterministic channel opening protocol for increasing values of $G$. Parameter $x$ is constant here, $x=0.05$.
    We present here the opening of 5 channels with conductance up to 4$G_0$.
The piecewise functions are presented in Appendix \ref{sec:App0}.
    }
    \label{fig:DCOP-tau}
\end{figure}

\begin{figure*}[htbp]
    \centering
    \hspace{-10mm}
    \includegraphics[width=1.1\linewidth]{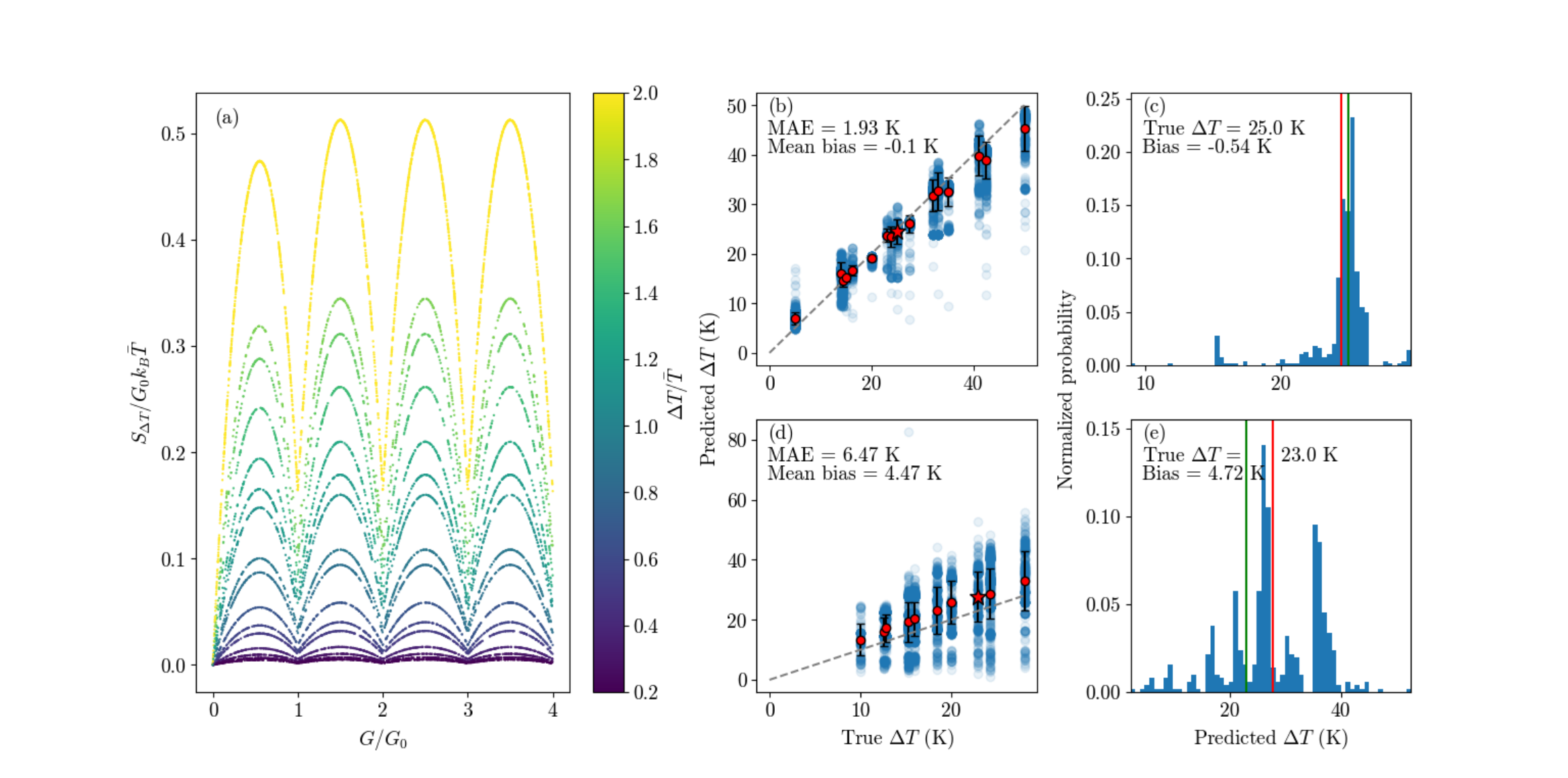}
    \caption{(a) Visualization of synthetic dataset generated by the deterministic channel-opening protocol with $G\leq4G_0$ using $x=0.05$. (b)-(e) NN predictions of $\Delta T$ values against true $\Delta T$, with example histograms of predicted values for one example of $\Delta T$. (b) NN predicted $\Delta T$ (blue) against true values on the synthetic training dataset, with the means of the distribution
    shown as the red circles and standard deviation as error bars. (c) Histogram displaying the distribution of predictions for a specific $\Delta T$, marked by the red star in (b). The green vertical line stands for the true $\Delta T$. The red vertical line indicates the mean of the predicted values in the histogram. (d) NN predicted $\Delta T$ against true values on the experimental dataset, with a corresponding histogram in (e).
    }
    \label{fig:DCOP}
\end{figure*}

\section{Synthetic data: Protocols}
\label{sec:Syn}
\subsection{Deterministic channel-opening protocol}

As a first attempt at generating synthetic data to provide data points in a wider range of $\bar{T}$ and $\Delta T$ values, we developed a procedure for computing $S_{\Delta T}$ given input values of $G$, $\Delta T$ and $\bar{T}$. Experimentally, these three quantities alone do not uniquely determine $S_{\Delta T}$, as is clear from observing the experimental data presented in Fig. \ref{fig:exp}. This is because $G$ sets only the sum of the transmission probabilities $\tau_i$ associated with all channels, without stipulating the exact value of each $\tau_i$.

To fill this gap, we follow a protocol for determining a typical value for each $\tau_i$ given the total conductance $G$, motivated by previous studies \cite{ruitenbeek1999}. This protocol requires a parameter be set, denoted by $x$, representing the rate at which transmission channels begin to partially open with increasing $G$, before all lower channels are completely open. $x$ can take any value between 0 and 1/2.

Without loss of generality, channel transmissions are indexed in decreasing order. For $0<G\leq 1G_0$, we assume that exactly two channels contribute to the overall conductance. By definition, $x=\tau_2G_0/G = \tau_2/(\tau_1 + \tau_2)$. Therefore, up to the conductance quantum, $\tau_2$ increases linearly with $G$ with a slope $x$. So, too, does $\tau_1$, with a slope $1-x$ as required for consistency. For $G>1G_0$, there are exactly three partially open channels at any time. As $G$ increases past $1G_0$, $\tau_3$ begins to grow as $\tau_3=x(G-G_0)$, the slope of $\tau_2$ increases to $1-2x$, 
and the slope of $\tau_1$ drops to $x$ so it can reach the fully open status in time for $G$ to reach $2G_0$, triggering the opening of channel 4. Once a channel is fully open, it remains so for all higher $G$ values. Appendix \ref{sec:App0} includes these piecewise functions, and Fig. \ref{fig:DCOP-tau} presents the protocol up to 4$G_0$.

Equipped with this protocol, we generated a synthetic dataset by first choosing a set of $\bar{T},\Delta T$ pairs. To address the false correlation seen between these quantities in the experimental dataset, we chose a grid-like array of points in the $\bar{T},\Delta T$ plane, as shown in Fig. \ref{fig:deltaT-T}. There, five $\bar{T}$ values were chosen at even intervals between 10 K and 25 K, reflecting the experimental values. At each $\bar{T}$, a range of $\Delta T$ values from 5 K up to the maximum possible, $2\bar{T}$ were used.

At each $\bar{T}, \Delta T$ pair, we generated 500 data points. For each, we chose the value of $G$ from a uniform distribution from 0 up to $1G_0$, $2G_0$, or $4G_0$. We then used our procedure to determine each $\tau_i$, and, in turn, $S_{\Delta T}$, via Eq.~(\ref{eq:S}). We chose to generate many conductance-noise pairs for a fixed set of $\bar{T},\Delta T$ values, rather than to choose a new random $\bar{T}$ and $\Delta T$ value every time we generate a point, to better mimic the experimental procedure. The resulting dataset is shown in Fig. \ref{fig:DCOP}(a) with $4G_0$ as the chosen upper limit to the range of $G$ values.

The dataset exhibits a periodic behavior of $S_{\Delta T}$ relative to $G$, particularly for $G>1G_0$, with slightly lower delta-T noise for $G<1G_0$ due to the presence of only two partially open channels. The quantity plotted is a non-dimensionalized version of the noise, $S_{\Delta T}/G_0k_B\bar{T}$, thus it depends on temperature only via a quadratic dependence on the ratio $\Delta T/\bar{T}$, leading to the observed stratification. This periodicity is due to the fact that only partially open channels contribute to the noise; for each interval between consecutive integer multiples of $G_0$, there are three partially-open channels opening in the exact same manner. Lower-indexed fully open channels contribute to $G$ as needed, but have no impact on $S_{\Delta T}$. This is a clear deviation from the experimental dataset, which shows a general trend towards overall growth of the delta-T noise with increasing $G$, see Fig. \ref{fig:exp}.

Furthermore, it is clear from the way the experimental data is distributed that knowledge of the quantities $G$, $\bar{T}$, and $\Delta T$ is insufficient, in practice, to determine the value of $S_{\Delta T}$, while in the synthetic data generated according to the deterministic protocol, it is.


In light of these significant qualitative differences between the synthetic and experimental datasets, 
we set out to develop a more complex protocol for generating synthetic data, with hopes of better capturing the physical processes at play.
Furthermore, as we show in Fig. \ref{fig:DCOP}(b)-(e) and discuss below in Sec. \ref{sec:ML}, 
while we successfully trained a model on deterministic synthetic data, the trained model, and thus the underlying synthetic data, were not able to capture experimental results. This failure demonstrates that the {\it deterministic} channel opening protocol, as depicted in Fig. \ref{fig:DCOP-tau}, does not represent the experimental situation.

\subsection{Noisy channel-opening protocol}

The experimental data (Fig. \ref{fig:exp}) differs significantly from the synthetic data generated according to a deterministic channel opening protocol (Fig. \ref{fig:DCOP}(a)), as seen most clearly by observing the scatter plots of $S_{\Delta T}$ vs $G$. For the deterministic data, the value of $S_{\Delta T}$ is determined by the values of $G$ and $\Delta T/\bar{T}$, without any variation. In contrast, the experimental data show a significant amount of noise, with a range of possible values for $S_{\Delta T}$ given $G$ at a particular $\Delta T/\bar{T}$. Further, the deterministic data exhibit periodic behavior, reaching minima at each integer value of $G/1G_0$. This is because the noise contributions of channels die off as they reach a fully open state. In contrast, in experimental data outside the $G<1G_0$ regime  (where quadratic dependence of $S_{\Delta T}$ on $G$ can be inferred with noise), there seems to be an accumulation of delta-T noise as $G$ increases, without observable minima of $S_{\Delta T}$ at integer values of $G/G_0$ higher than 1.

We set out to develop a procedure for generating synthetic data that more accurately reflects the physics of channel opening in nanoscale junctions, as reported in experiments \cite{Vardimon2013} and calculations \cite{Stafford,pauly2011,Cuevas2017}, leading to a dataset that better reflects experimental noise observations. We specifically found that emulating the channel opening trends seen in Ref. \citenum{pauly2011} provided good description of the experimental results. As such, we built on computations from this paper, along with Ref. \citenum{Vardimon2013}, to guide us in generating synthetic datasets. 

However, we point out that these studies considered gold nanojunctions without introducing hydrogen molecules into the contact region, in contrast to the experimental data we try to capture. Some of the hydrogen molecules are likely to settle in the small atomic gap formed between the electrodes (leading to conductance values below 1$G_0$), while others penetrate the gold contact region and the bulk affecting transport even at higher conductance. Therefore, the presence of hydrogen molecules probably affects the conductance in the range up to 4$G_0$, which we focus on. As such, Refs. \cite{Vardimon2013,pauly2011} only serve us as a general guide for the opening of channels in gold nanojunctions. 

In the deterministic protocol, one single parameter is set, namely, $x$, dictating the rate at which partially open, higher-indexed channels open up; see Fig. \ref{fig:DCOP-tau}.  
For the noisy protocol, we augment this by several ingredients: (S1) Sampling $x$ from a probability distribution, so it takes on a different value in generating each data point. 
(S2) 
Increasing the number of partially open channels with increasing conductance. This is done by
limiting the transmission of channels above channel 1 to no more than 0.95 and defining additional fixed parameters for the different rates of change of $\tau_i$ for $i>2$. 
 (S3) Adding some noise to the resulting distribution of $\tau_i$ that builds a given $G$.
There is some arbitrariness in our parametrization in step (S2), which is handled by this randomization. 

\begin{figure}
    \centering
    \includegraphics[width=1\linewidth]{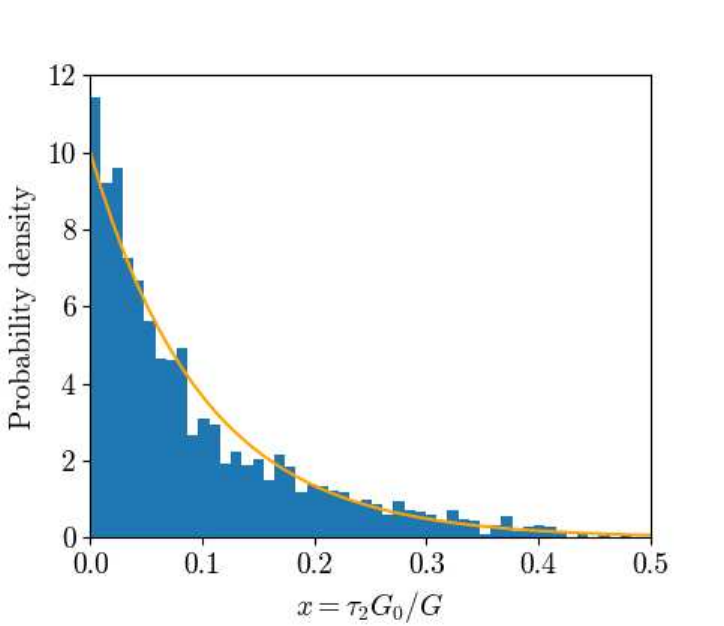}
    \caption{Histogram of $x$ values obtained from experimental data, scaled as a normalized probability density function. We acquire the different values of $x$ by assuming that only two channels have nonzero transmission for conductance less than $1 G_0$. The orange curve is an exponential distribution characterized by the mean value of the experimental $x$ values.}
    \label{fig:x_dist}
\end{figure}

We now explain in more detail the elements in the generation of noisy synthetic data.

(S1) 
To generate each data point, the value of $x$ is chosen by sampling an exponential distribution characterized by a mean value of $0.1$. We choose this distribution by looking at the experimental data for $G<1G_0$, where we assume that at most two channels have non-negligible transmission. As such, the measured values of $G$ and $S_{\Delta T}$, given $\bar{T}$ and $\Delta T$, are sufficient to calculate the values of $\tau_1$ and $\tau_2$, and, in turn, $x$, see the resulting scaled histogram in Fig. \ref{fig:x_dist}. Note that the highest possible value of $x$ is 0.5 since we choose to label the channel with a higher transmission value with the index $1$. 

The exponential distribution was chosen as it reflects the experimental behavior quite well while also being defined by only a single parameter (its mean), maintaining some simplicity. Technically, this distribution is nonzero at all positive values. As such, in the protocol, we choose to reject values higher than 0.4. Although values as high as 0.5 are physically possible, we find that they are suppressed beyond what is captured by the exponential distribution, as seen in Fig. \ref{fig:x_dist}. The mean value of $x$ determined in this way from the experimental data is 0.095. 
For convenience, we round this to 0.1 in generating the synthetic data, as we expect other sources of uncertainty (e.g., rejection of values above 0.4) to render this distinction insignificant.

\begin{figure*}[htbp]
    \centering
    \includegraphics[width=0.85\linewidth]{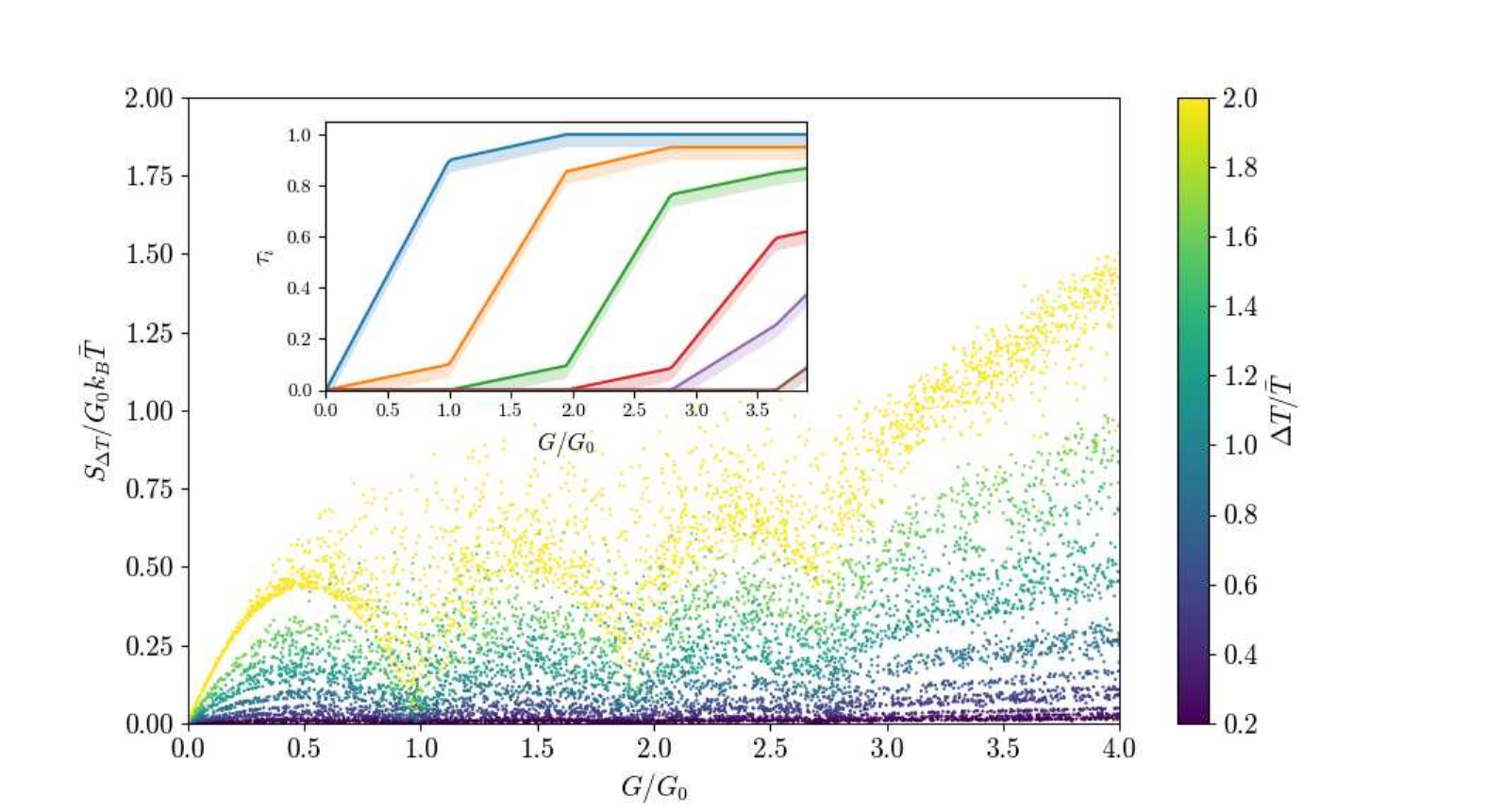}
    \caption{Synthetic datasets with $0\leq G\leq4G_0$, as generated using the noisy channel opening protocol. Each data point is generated using a value of $x$ sampled from the exponential distribution whose mean is 0.1 (with only values up to 0.4 accepted) and additional parameters characterizing the growth of channel transmissions at higher $G$ values are chosen. Values of each $\tau_i$ determined according to the procedure described in the main text. Inset: the transmission probability, $\tau_i$, of successively opening channels (i.e., $i$ increases from left to right along the plot) given a value of $G$. Each channel opens accordingly to a piecewise defined function of $G$ mimicking previous studies \cite{Vardimon2013, pauly2011}. Shaded regions represented additional noise added onto transmissions to capture additional sources of variation.}
    \label{fig:noisy_synthetic_data}
\end{figure*}

(S2) In addition to choosing $x$, we choose a value of $G$ by sampling a uniform distribution between 0 and a set maximum value (usually $G_0$, $2G_0$, or $4G_0$). Given this value, the transmission probability associated with each channel is set according to the predetermined piecewise defined functions for the $\tau_i$'s in terms of $G$, plotted in the inset of Fig. \ref{fig:noisy_synthetic_data}. 
For details, see Appendix \ref{sec:App0}.
%

(S3)  For each data point, we introduce additional noise by taking the value of each $\tau_i$ determined by the functions generated in Appendix \ref{sec:App0} and subtracting a value sampled from a uniform distribution between 0 and 0.05, provided that doing so does not result in a negative value. This accounts for the fact that the channels are not expected to open in the exact same manner for every run of the experiment. The shaded regions in the inset of Fig. \ref{fig:noisy_synthetic_data} hence represent the range of possible values for each $\tau_i$ given $G$. We note that this addition of noise will change the sum of the $\tau_i$'s, thus, the value of $G$ must be recalculated as it will differ slightly from the input value initially used to calculate the transmission probabilities.

These three changes introduce variations to the possible value of each $\tau_i$ given the value of $G$, manifesting as variations in $S_{\Delta T}$. 
Crucially, preventing each channel from reaching a transmission probability of 1 at periodic steps in $G$ leaves more channels with a nonzero contribution to the noise that persists as $G$ grows, exhibiting the experimentally observed trend of higher noise at higher conductance, see Fig. \ref{fig:noisy_synthetic_data}.



The nonequilibrium-excess noise $S_{\Delta T}$ is then calculated according to Eq.~(\ref{eq:S}). Results are shown in Fig. \ref{fig:noisy_synthetic_data}. The periodic behavior seen with the deterministic protocol has been significantly obscured here, and a trend of increasing $S_{\Delta T}$ with $G$ is now present, comparable to the experimental data.
However, minima at multiples of $G_0$ can still be inferred, though they become less pronounced with increasing conductance. This reflects that the uncertainty in the manner in which channels are opening grows and more channels are at play.

Another notable feature of the synthetic datasets is the general trend for $S_{\Delta T}$ to increase with $\Delta T/\bar{T}$, although the noise makes deviations from this trend possible. The apparent stratification of datasets is due to the fact that only a discrete set of $\Delta T/\bar{T}$ values were used, rather than a continuous range. This behavior is expected based on the theoretical expression for $S_{\Delta T}$. No such trend can be observed in the available experimental data, but this is simply because the experimental datasets do not cover a wide enough range of $\Delta T/\bar{T}$ values, as discussed above.


\section{Training and Testing}
\label{sec:ML}

\subsection{Procedure and Metrics}
We build a deep learning model with the Keras library \cite{Keras} framework for the prediction of $\Delta T$. The model basis is a feedforward NN consisting of hidden layer(s) between input and output layers. The network architecture, such as a loss function and an optimizer, as well as the performance evaluation to choose the number of layers and neurons, is described in Appendix \ref{sec:App1}.

In accordance with our objective, models were trained on scaled features $G/G_0$, $S_{\Delta T}/G_0k_B\bar{T}$, and $\bar{T}$ (input layer) from synthetic datasets to target corresponding $\Delta T/\bar{T}$ (output layer),  which we scale back to $\Delta T$ as the final predicted value. Each synthetic dataset contains 12,500 total data points, divided to 625 points for each of the 20 ($\bar{T}$, $\Delta T$) pairs. The model is trained on 10,000 points (80\% of dataset - 20\% for testing). This total number of data points remains constant when we vary the maximum $G$ allowed in the dataset. When using a trained model to predict on the experimental dataset, we limit that set to the same maximum $G$. Overall, the nature of predictions is that a distribution of predicted $\Delta T/\bar{T}$ is formed for a given true $\Delta T$. This is the result of the nonunique and non-deterministic (noisy) relation between the current noise $S_{\Delta T}$ and the conductance, and $\Delta T/\bar T$.

Two metrics are used for the evaluation: the mean absolute error (MAE) between any predicted $\Delta T$ and true $\Delta T$ values, and the mean bias, defined as the difference between the predicted and true $\Delta T$, averaged over the entire dataset--that is the mean {\it signed} error. 
A positive mean bias value indicates that the predicted mean is larger than the true one, and vice versa. 
In histogram visualizations, we also use ``bias" to quantify the signed $\Delta T$ difference between the mean of predictions in the ensemble and the true $\Delta T$, for a unique value of $\Delta T$. 
We use the trained model to predict on its own training dataset, as well as predict on the experimental dataset. We also show examples of histograms of a prediction distribution for a given true $\Delta T$.

\subsection{Failure of the deterministic protocol}

We described in Sec. \ref{sec:Syn} the generation of synthetic datasets according to the simple fixed-$x$ rule, depicted in Fig. \ref{fig:DCOP-tau}, with noise-conductance results presented in Fig. \ref{fig:DCOP}(a). We continue and discuss in Fig. \ref{fig:DCOP}(b)-(e) the predicted results of the NN when trained on this synthetic dataset of the deterministic channel opening protocol.

Fig. \ref{fig:DCOP}(b)-(c) presents testing on synthetic data.
The metrics indicate that the MAE is 1.93 K and the mean bias is -0.1 K, that is, the majority of distribution means were less than the true value, and within 1 K in proximity. 
An example histogram is shown in panel (c) for a true value $\Delta T=25$ K (marked by a star in (b)). 
We find that the center of the main distribution lies almost directly on the true $\Delta T$ (green line). However, the distribution mean (red line) is smaller due to outliers predicted much below the majority of results. This difference results in a bias of -0.54 K. 

We proceed and employ this trained model to predict on features in the experimental dataset.
Results are presented in Fig. \ref{fig:DCOP}(d) showing mean absolute errors of about 6.5 K and a mean bias of about 4.5 K.
Looking at an example histogram in Fig. \ref{fig:DCOP}(e), significant number of predictions appear around $\sim38$ K, whereas the true value is $23$ K, leading to
a distribution mean that is greater than the true value. 
Since training and testing on the synthetic data was successful, as demonstrated in 
panels (b)-(c), it is evident that the trained model failed to capture experiments due to the synthetic data incorrectly representing experimental trends. 


\begin{figure*}[htbp]
    \centering
    \hspace{-10mm}
    \includegraphics[width=1.1\linewidth]{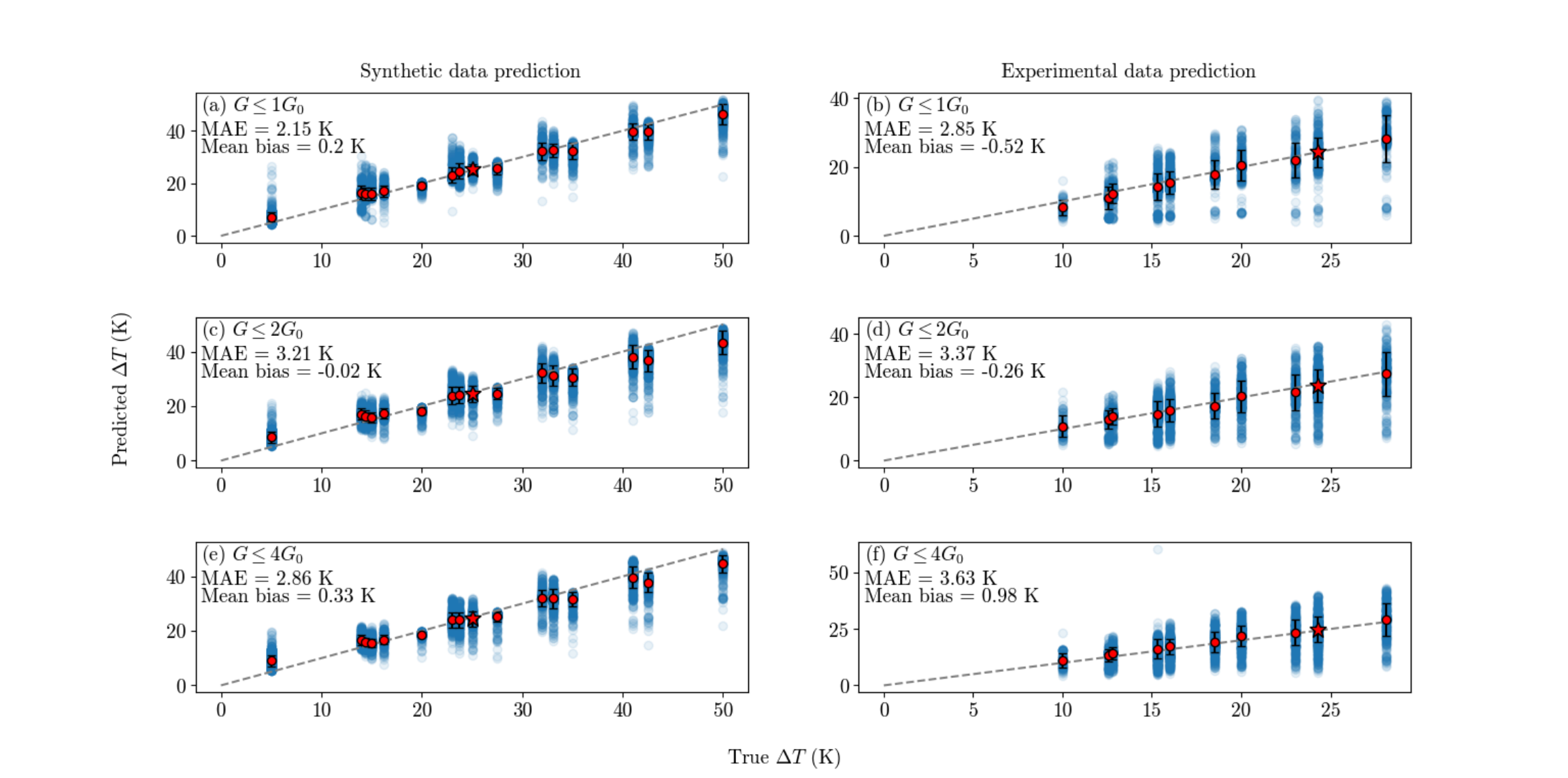}
    \caption{Neural network predicted $\Delta T$ plotted against true $\Delta T$ with model trained from datasets generated by the noisy channel opening protocol. Left column ((a), (c), (e)) shows predictions on the synthetic training dataset; right column ((b), (d), (f)) shows predictions on the experimental dataset. Rows are separated by maximum $G$ allowed in all datasets.
    }
    \label{fig:NCOP-TvT}
\end{figure*}

\begin{figure*}[htbp]
    \centering
    \hspace{-10mm}
    \includegraphics[width=1.1\linewidth]{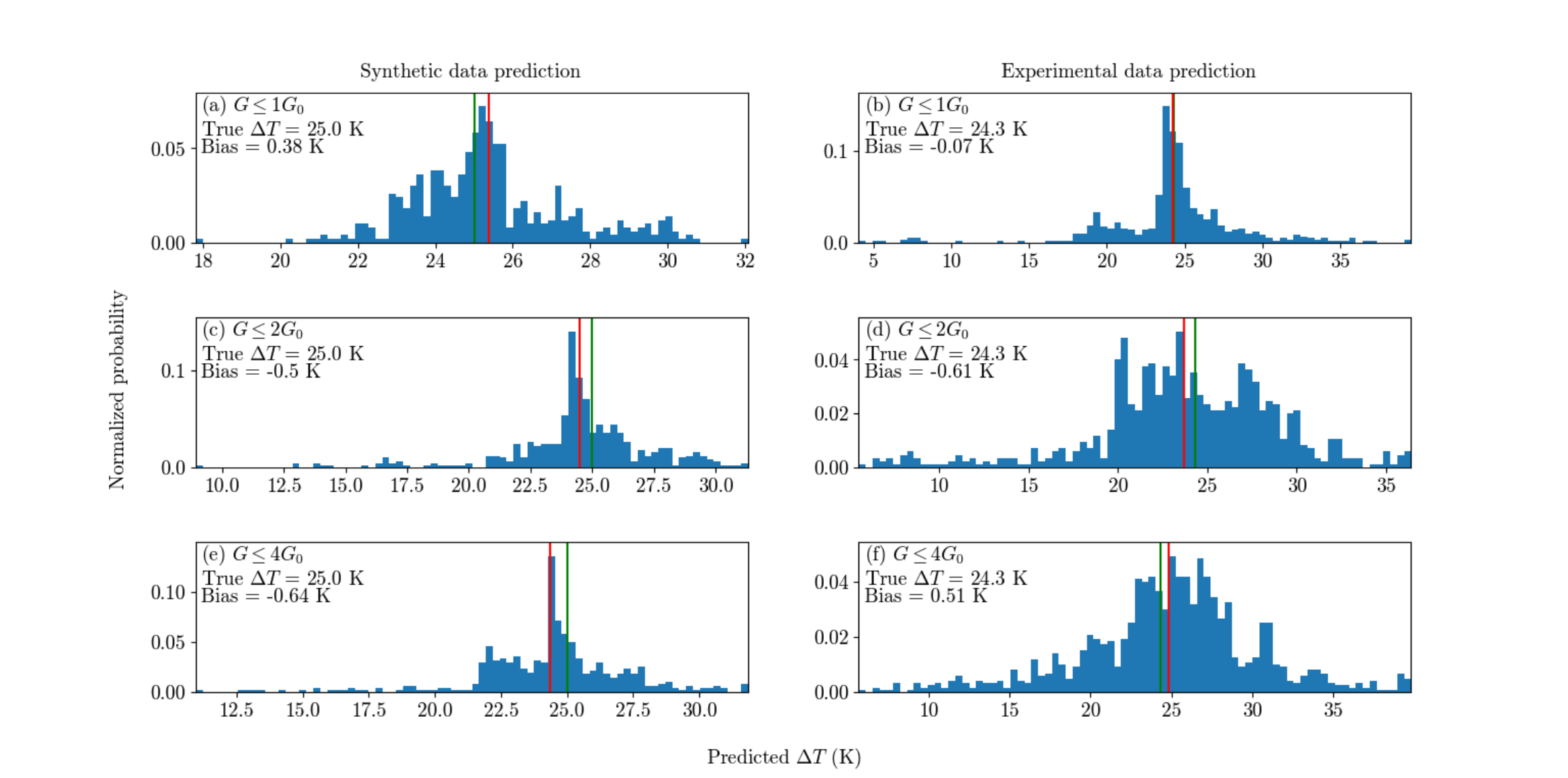}
    \caption{Histograms of single chosen true $\Delta T$ from corresponding panels in Fig. \ref{fig:NCOP-TvT} as indicated there by the red stars. The true $\Delta T$ is shown within plot by the green vertical line, and the mean of the histogram is shown by the red line. True temperature data is consistent in columns: Left column ((a), (c), (e)) share $\bar{T}=17.5$ K and true $\Delta T=25.0$ K; right column ((b), (d), (f)) share $\bar{T}=21.5$ K and true $\Delta T=24.3$ K.}
    \label{fig:NCOP-hist}
\end{figure*}
\vspace{10mm}


\begin{figure*}[htbp]
    \centering
    \hspace{-10mm}
    \includegraphics[width=1.1\linewidth]{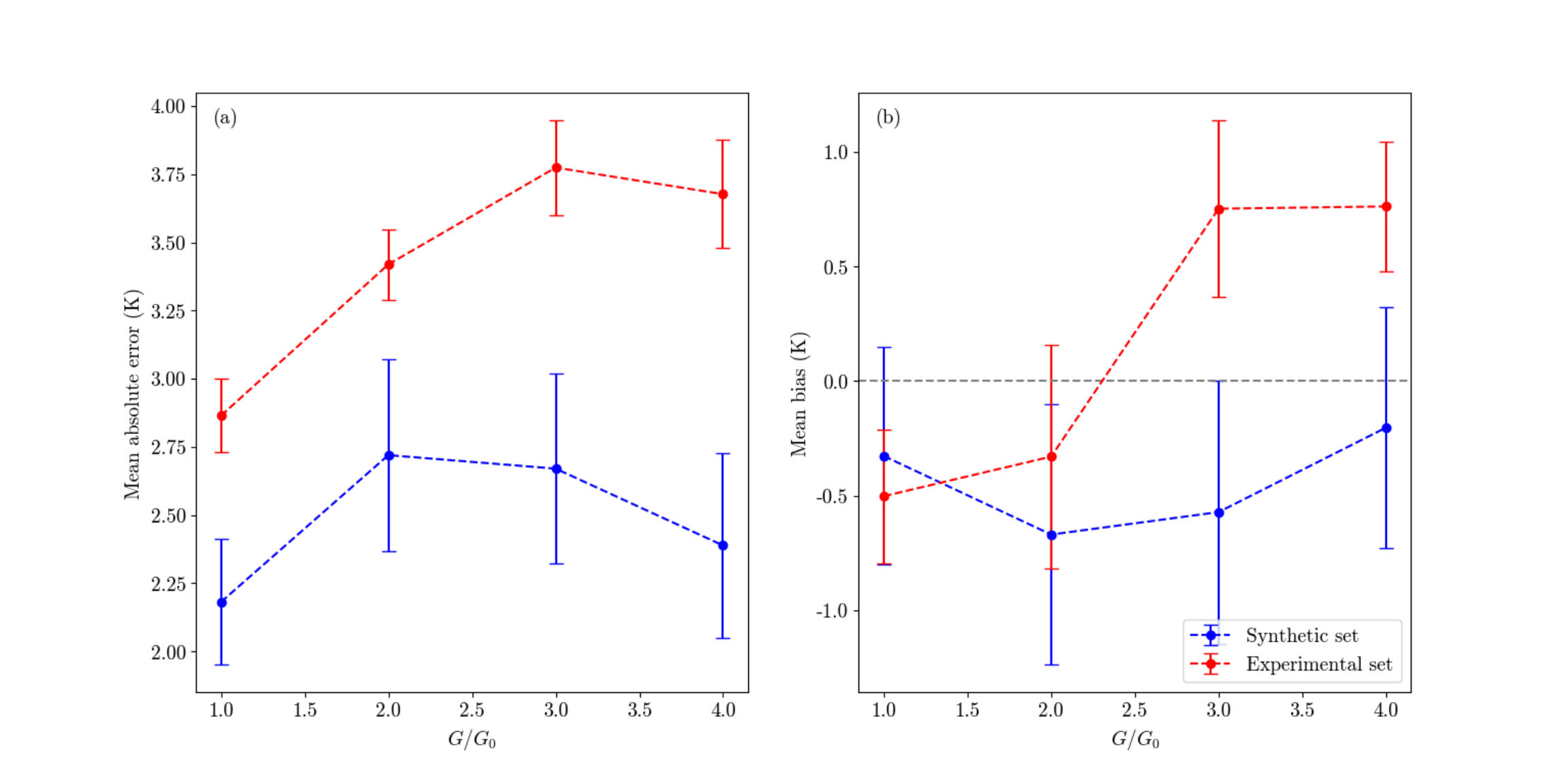}
    \caption{Metrics of (a) mean absolute error and (b) mean bias of neural network $\Delta T$ predictions with varying max $G$ allowed in datasets. Each data point is averaged over 10 retrained models, and error bars display the standard deviation in the models. Data in blue indicates predictions on synthetic (training) dataset and data in red indicates predictions on experimental dataset.}
    \label{fig:NCOP-G}
\end{figure*}
\subsection{Training and testing data from the noisy protocol}

We now turn our attention to deep learning models trained from datasets generated via the noisy channel opening protocol. 
As we show in Figs. \ref{fig:NCOP-TvT}-\ref{fig:NCOP-G}, these models perform well when tested on experimental datasets, providing reasonable predictions of $\Delta T$ 
with mean biases smaller than 1 K. 
This good performance provides confidence in the channel opening protocol \cite{Vardimon2013,pauly2011}, supports the theoretical expression (\ref{eq:S}) --- and opens the door to the estimation of the temperature bias based on measurements of the delta-T shot noise. 

We now detail our results.
Fig. \ref{fig:NCOP-TvT} shows the predictions of the synthetic set (left) and the experimental set (right) against true $\Delta T$ for models trained with increasing maximum conductance $G$ (top to bottom) in the datasets.
When using the models to predict on its synthetic training set, Fig. \ref{fig:NCOP-TvT}(a), (c) and (e), a notable observation is that for some values, e.g. $\Delta T=20 $ K and $\Delta T=50$ K, the predictions are (i) relatively narrow in the distributions, i.e. precise, and (ii) there are only a few predicted values exceeding true value. This is because those datasets were generated
for the maximally allowed $\Delta T = 2\bar{T}$, using $\bar T=10$ and $\Delta T=20$ K, see Fig. \ref{fig:deltaT-T}. 
Similarly, predictions at our lower temperature bias $\Delta T=5$ K are mainly higher than the true value.
These observations reflect that the neural network has learned the range of allowed $\Delta T$. 

Predictions on the experimental dataset are presented in Fig. \ref{fig:NCOP-TvT}(b), (d) and (f). Here, we observe a decrease in the MAE, and, more notably, the mean bias down to within 1 K, compared to metrics shown for the deterministic datasets in Fig. \ref{fig:DCOP}(d). 
These improved metrics can also be seen from the distributions of predicted $\Delta T$,
Fig. \ref{fig:NCOP-hist}. In each histogram we observe that the distribution peak and thus the mean (in red) is in proximity to the true $\Delta T$ (in green). While these histograms are wide, they are generally evenly distributed above and below the true $\Delta T$. 

In Fig. \ref{fig:NCOP-G}, we compile our neural network metrics with increasing max $G$ in datasets. Metrics are averaged over 10 models. 
We generally find MAE up to 3.75 K and mean biases in the range of -0.5 to 1 K.
In this respect, it is useful to comment that the experimental data has an uncertainty of about 0.5 K in $\Delta T$. We therefore regard models with mean biases of up to 1 K as ``predictive".
We make the following observations:
(i) 
Predictions on experimental datasets (red) suffer from higher errors (MAE, mean bias) compared to predictions over synthetic sets (blue).
(ii) Predictions on experiments are typically more accurate for small $G$, with error metrics growing with $G$. 

Both of these observations can be rationalized by recalling that training was done on synthetic-noisy data, with a protocol for channel opening inspired by computations \cite{pauly2011} and experiments \cite{Vardimon2013}.
Our protocol can only mimic general trends rather than the precise channel opening process. Thus, it is not surprising that error metrics are larger when models that are trained on synthetic data are tested on experimental sets. Similarly, our protocol for channel opening appears to be closer to reality when only a few channels are involved. 
Note that experimental datasets include junctions with conductance exceeding 4$G_0$, see Fig. \ref{fig:exp}. However, in order to make $\Delta T$ predictions about these high-conductance junctions, synthetic data are needed in that range, making more assumptions about channel openings. Given the scarcity of studies on channel opening in relevant junctions at $G>4G_0$, we limited our predictions below that range. 

In Appendix \ref{sec:App2}, we generate synthetic datasets based on an integral formula from which Eq. (\ref{eq:S}) was derived after additional approximations.
Since training and testing on this synthetic dataset yield results comparable in quality to those based on the approximate Eq. (\ref{eq:S}), we conclude that this approximate equation provides a valid and practical alternative to the more accurate yet cumbersome expression [Eq. (\ref{eq:Sor})], and that errors in our predictions on experiments do not stem from the use of the approximate expression.

\begin{figure}[htpb]
    \centering
    \includegraphics[width=1\linewidth]{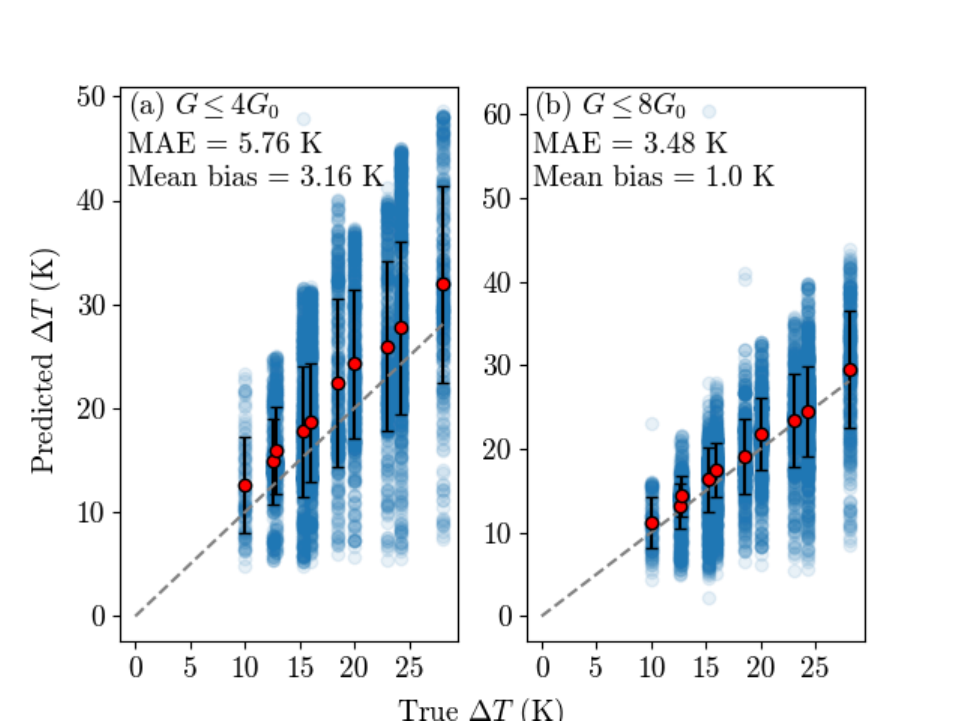}
    \caption{Extrapolation tasks. Neural network predictions of $\Delta T$ on experimental data. We use a model trained with synthetic data on   lower $G$ regimes for extrapolations on experiments to higher $G$ values. (a) Predictions from a model trained on synthetic data with $G \leq 2G_0$, applied to experimental data with conductance up to $4G_0$. (b) Predictions from a model trained on synthetic data with $G \leq 4G_0$, applied to experimental data with conductance up to $8G_0$.}
    \label{fig:G_extra}
\end{figure}
\subsection{Extrapolation tasks}
What about extrapolation tasks? 
We consider a scenario where datasets are collected from junctions with relatively low conductance, and we are tasked with predicting $\Delta T$ for junctions with potentially higher conductance. As we show in Fig. \ref{fig:G_extra}, this type of extrapolation can be achieved, depending on the training set. In particular, this can be done without any modifications to our neural network architecture or training procedure.

In Fig. \ref{fig:G_extra}(a), we train a model using noisy synthetic datasets with conductance up to $G\leq2G_0$, and then apply it to predict on experimental data with higher conductance, up to $4G_0$. The performance metrics are suboptimal, 
particularly when compared with predictions made within the same range of $G$ on which the NN was trained (recall Fig. \ref{fig:NCOP-TvT}(d)). 
In contrast, in Fig. \ref{fig:G_extra}(b) we apply a model trained with $G\leq4G_0$ to essentially all available experimental data, with $G$ ranging up to $8G_0$. In this case, the NN performs significantly better, with metrics comparable to those shown in Fig. \ref{fig:NCOP-TvT}(f). 

We can rationalize these results as follows. When trained exclusively on data with low conductance, the NN fails to learn that with higher conductance, more transmission channels should open sequentially and partially. As a result, instead of ``opening" more channels to explain the elevated noise at higher conductance, the NN predicts higher temperature biases in Fig. \ref{fig:G_extra}(a).
In contrast, when the model is trained on data up to $G \leq 4G_0$, where up to 6 channels are involved, as seen in Fig. \ref{fig:noisy_synthetic_data}, the neural network successfully captures the increase in noise caused by the sequential opening of channels. This enables the model to make more accurate predictions, as shown in Fig. \ref{fig:G_extra}(b).

However, challenges arise when attempting to extrapolate $\Delta T$ over a wider temperature range. Specifically, when models are trained on datasets with low $\Delta T$ and tasked with predicting higher $\Delta T$ values, outside the training range, we observe that the neural network ``resists" making predictions beyond its learned domain.  A similar rigidity occurs when training on high $\Delta T$ datasets and attempting to predict lower $\Delta T$.
To improve the model's performance in extrapolation tasks, more efforts should be placed on the NN architecture and training strategies, a direction that we leave for future work.

\section{Summary}
\label{sec:Summ}

Noise can serve as a probe to infer the properties of the system and the underlying transport mechanisms. In this work, we demonstrated that delta-T noise, combined with conductance measurements, can be used to estimate temperature biases in atomic-scale junctions. We accomplished this task by employing a supervised machine learning algorithm, training models on delta-T shot noise, conductance, and average temperature to predict $\Delta T$. 

Due to the limited range of $\bar T$, $\Delta T$ parameters in the available experimental data, we generated synthetic datasets emulating single-molecule transport experiments. A key conceptual challenge in this process was to capture the dependence of the noise and conductance on the transmission probabilities of multiple channels, which are determined by microscopic details not directly accessible through experiments.
We tested two data-generation protocols: a deterministic approach and a noisy channel-opening method. Although the deterministic approach yielded results inconsistent with experiments, models trained on noisy datasets produced reasonably accurate predictions when applied to experimental data, with comparable prediction errors across both training and testing datasets.

Our study contributed to both practical and fundamental aspects. On the practical side, we demonstrated that stimuli controlling transport in atomic-scale junctions can be effectively extracted from noise measurements by using supervised ML methods.
From a fundamental perspective, we found that training models on synthetic data, which remain consistent with experimental observations, enables us to:
(i) Gain physical insight into the system, particularly trends associated with channel openings. 
(ii) Support the analytical delta-T formula in a broader conductance range, beyond 1$G_0$, which had previously been studied in Ref. \citenum{DeltaTShot}.
(iii) Support the analytical delta-T formula in a broader range of temperatures, even in regimes where $\Delta T$ approaches $2\bar{T}$---a scenario where the temperature of one terminal becomes very low, approaching zero.



The principle behind our supervised ML approach for predicting $\Delta T$ from delta-T shot noise can be applied similarly to predict temperature biases using the more commonly available delta-T flicker noise \cite{flicker2}. More broadly, our workflow can be used to estimate other stimuli or internal parameters from molecular junction measurements. 

Measurement of temperature differences at the nanoscale is essential for advancing understanding of thermal transport in nanodevices. Accurate temperature measurements can, for example, drive the development of thermoelectric devices, improve thermal management of electronic devices, and advance the performance of quantum information technologies. Noise signals can be a powerful tool for probing local thermal biases. With ongoing progress in ML methods, we envision these techniques becoming a more standard part of experimental workflows, allowing a rapid estimation of parameters and assessment and verification of measured results. 




\section*{Acknowledgments}
We acknowledge discussions with Matthew Pocrnic and Juan Felipe Carrasquilla. The work of MG and JJW was supported by the NSERC Canada Graduate Scholarship-Doctoral. 
DS acknowledges support from an NSERC Discovery Grant and the Canada Research Chair program. 
OT acknowledges the support of the European Research Council (Grant 864008), the Israel Science Foundation (Grant No. 2129/23), funding by the Harold Perlman family, and research grants from Dana and Yossie Hollander.

\appendix

\section{Details on channel opening protocols}
\label{sec:App0}

We elaborate here on our construction of the piecewise functions that build the total transmission, $\tau=\sum_i \tau_i$, with $G=G_0\tau$.
In the deterministic protocol, we list here the piecewise functions in several regimes with $x<1$ dictating the rate of channel opening; see Fig. \ref{fig:DCOP-tau}.

For $0\leq\tau\leq1$,
\bea
\tau_1= (1-x)\tau,\,\,\,\,\,
\tau_2=x\tau.\,\,\,\ 
\eea
For $ 1\leq\tau\leq2$, we define
$\tau_M\equiv\tau-1$,
\bea
\tau_1&=& (1-x) + x\tau_M,\,\,\,\,\,
\tau_2=x+(1-2x)\tau_M,\,\,\,\ 
\nonumber\\
\tau_3&=&x\tau_M. \,\,\,\,\
\eea
For $ 2\leq\tau\leq3$, we define
$\tau_M\equiv\tau-2$,
\bea
\tau_1&=&1,\,\,\,
\tau_2=(1-x) + x\tau_M,\,\,\,\,\,
\tau_3=x+(1-2x)\tau_M,\,\,\,\ 
\nonumber\\
\tau_4&=&x\tau_M,\,\,\,\,\
\eea
and continuing in this pattern to higher conductance.

In the noisy-opening protocol, we elaborate here on step (S2).
Recall that $x$ is sampled from an exponential distribution, and that it is the slope of each $\tau_i$ for channels 2-4 as they begin to open, up until their transmission probabilities reach a value of $x$. For consistency, up until $1G_0$, the slope of $\tau_1$ is $1-x$. After the first kink, for $1G_0<G\leq 1.95G_0$, $\tau_1$ has a slope of $(20/19)x$ until it saturates at 1, while $\tau_2$ grows with a slope of $1 - (39/19)x$. For $1.95G_0 < G \leq 2.8G_0$, while channel 4 is beginning to open with slope $x$, $\tau_3$ grows with a slope of $1 - (36/17)x$ and $\tau_2$ with a slope of $(19/17)x$. For $2.8G_0<G\leq 3.65G_0$, channels 1 and 2 are both saturated at transmissions of 1 and 0.95 respectively, while $\tau_3$ grows with a slope of $x$ and $\tau_4$ with a slope of 0.6, slower than channels 2 and 3 grew after their first respective kinks (based on a typical $x$ value of around 0.1). Channel 5 begins to open with a slope of $0.4-x$, which is usually greater than $x$. For $G>3.65G_0$, $\tau_3$'s growth slows to a slope of 0.07, $\tau_4$ grows with slope $x$, $\tau_5$ grows with slope $0.38-x$, and channel 6 opens, with its transmission growing at a slope of $0.35$. We find that six transmission channels are sufficient to cover the range of conductance values up to $4G_0$.

The protocol above is obviously somewhat arbitrary in that the slopes chosen and the residual opening before saturation could have been constructed with different numbers to mimic experiments. To relax these rules and get closer to the experiment, step (S3) as detailed in the main text adds some randomness to the channel opening protocol.  
Together, steps (S1)-(S3) generate data sets pictured in Fig. \ref{fig:noisy_synthetic_data}.

\begin{figure*}[htbp]
    \centering
    \includegraphics[width=1.05\linewidth]{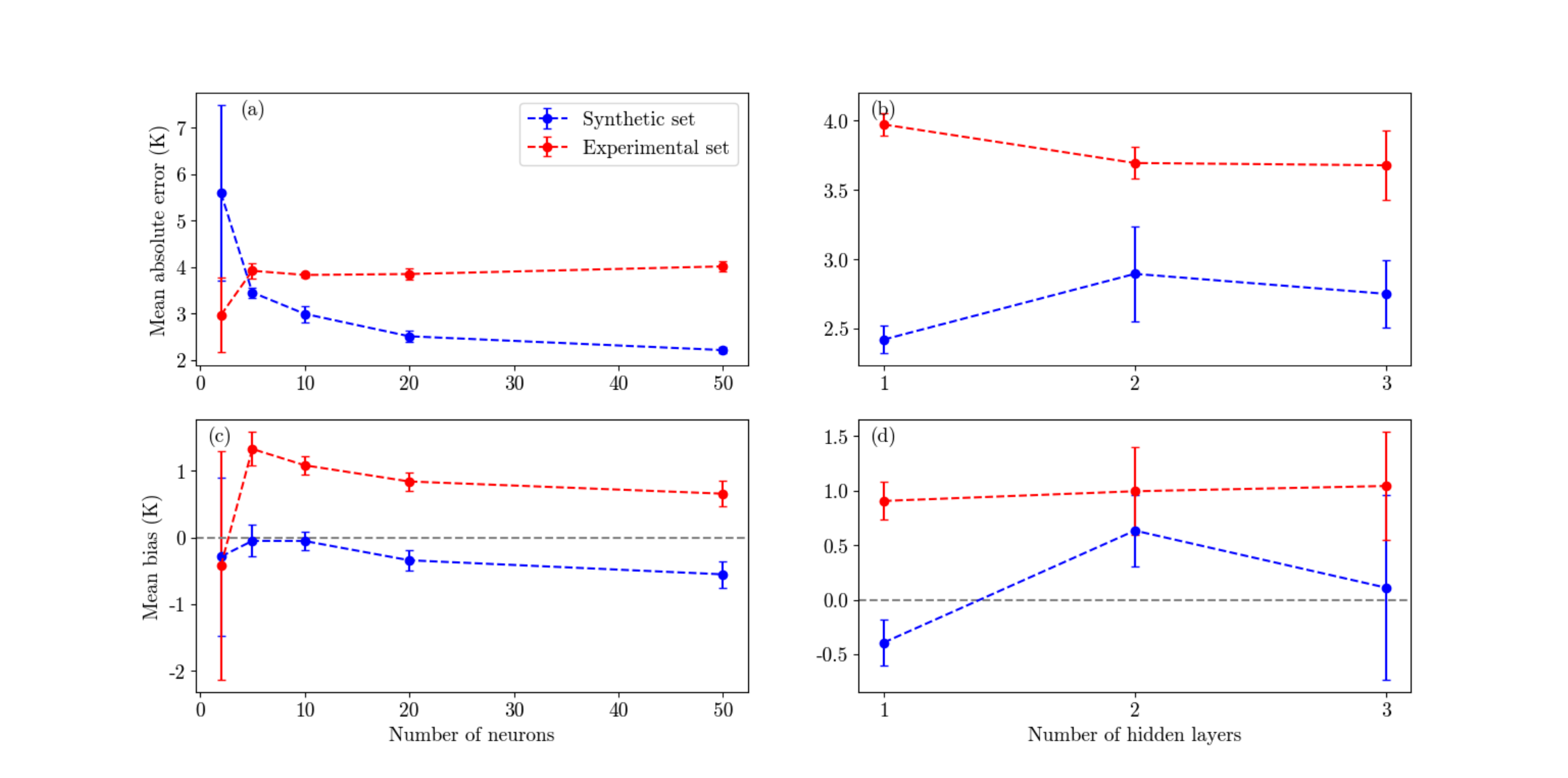}
    \caption{Neural network architecture: Performance evaluation in terms of the mean absolute error and mean bias metrics of deep learning models trained with various number of (a), (c) neurons in one hidden layer and (b), (d) hidden layers each with 20 neurons. Training was performed with synthetic dataset generated by the noisy channel opening protocol with conductance limited to $G\leq4G_0$. Each data point indicates the averaged MAE and mean bias over 10 trained models with the same setting, and error bars display the standard deviations in results from those 10 models.}
    \label{fig:NN-test}
\end{figure*}

\begin{figure*}[htbp]
    \centering
    \includegraphics[width=1.05\linewidth]{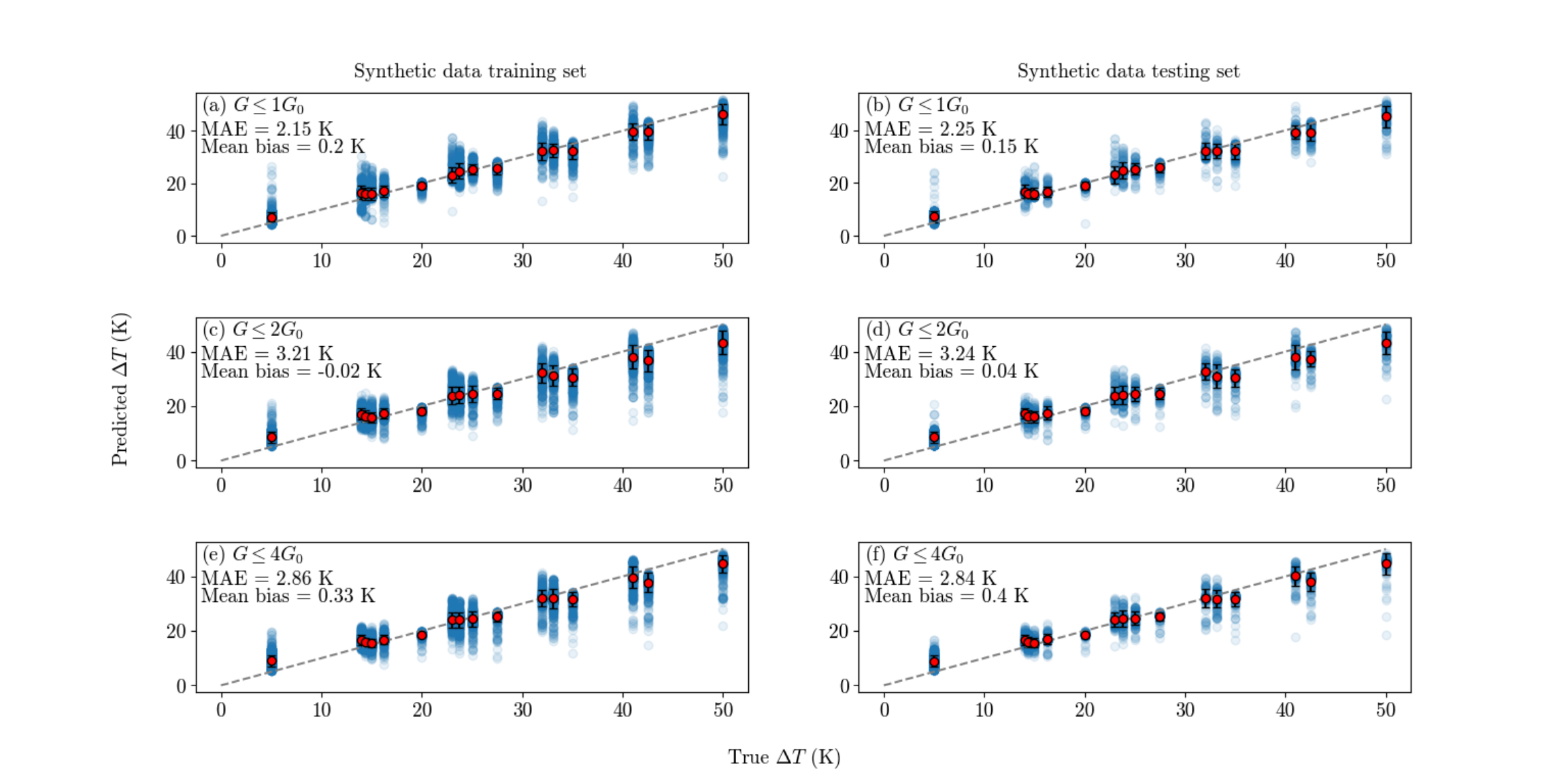}
    \caption{Training and testing the ML model on synthetic data. We present the NN predicted $\Delta T$  against true $\Delta T$. Datasets, generated with the noisy channel opening protocol, were split to 80\% training and 20\% testing sets. Left column panels (a), (c), (e) show predictions on the training set; right column panels (b), (d), (f) show predictions on the testing set. Each row presents results with datasets allowing increasingly higher maximum $G$.}
    \label{fig:NN-train-test}
\end{figure*}

\vspace{5mm}
\section{Neural network architecture}
\label{sec:App1}

Deep neural networks were constructed and trained using the Keras library \cite{Keras}. The general NN architecture is a feedforward network with a number of hidden layers and neurons chosen after performance evaluation, as shown in Fig. \ref{fig:NN-test}. 
From panel (a) we see that trained models predicting on their own training sets, or predicting on the experimental sets, show opposite trends for the MAE as the number of neurons increases: While predictions on synthetic sets are improved with a growing number of neurons, the opposite trend shows when testing on experimental data. We attribute this to some form of overfitting on training data, leading to worse predictions on unknown experimental data. 

Interestingly, with respect to the mean bias metrics, panel (c) shows predictions on synthetic and experiments having mean biases that show similar trends with increasing number of neurons at different magnitudes. Overall, the increase in MAE of the predictions from the experimental data is not as drastic as the improvement of the predictions from the training set. Achieving a low standard deviation in both MAE and mean bias, we choose 20 neurons in a hidden layer to be the standard setting of models for main results.

Regarding the number of hidden layers, Figs. \ref{fig:NN-test}(b) and (d) show minimal improvement in error metrics when increasing the number of hidden layers, with 20 neurons used in each layer. We eventually chose 3 layers as our main architecture. This allows balancing best metrics and low fluctuations for predictions on both training and experimental data. 

Thus, the overall NN architecture consists of an input layer, 3 hidden layers with 20 neurons each, and an output layer. All hidden layers include dropout (20\%) and use the rectified linear unit (ReLU) activation function. 
For training, we use Adam optimization \cite{Adam} on the mean absolute error (L1) loss function, $L = \frac{1}{N}\sum_{i=1}^{N} |y_{\text{true}} - y_{\text{i,pred}}|$, where $N=32$ is the batch size, 
and we iterate over 100 epochs with constant learning rate.

The performance of the model on the test set (20\% of the synthetic dataset) is shown in Fig. \ref{fig:NN-train-test}. We find that there is no significant overfitting on the training set as the resulting error metrics for predictions on the testing set are highly similar to metrics received on the training set.


\begin{figure*}[htbp]
    \centering
    \hspace{-15mm}
    \includegraphics[width=1.1\linewidth]{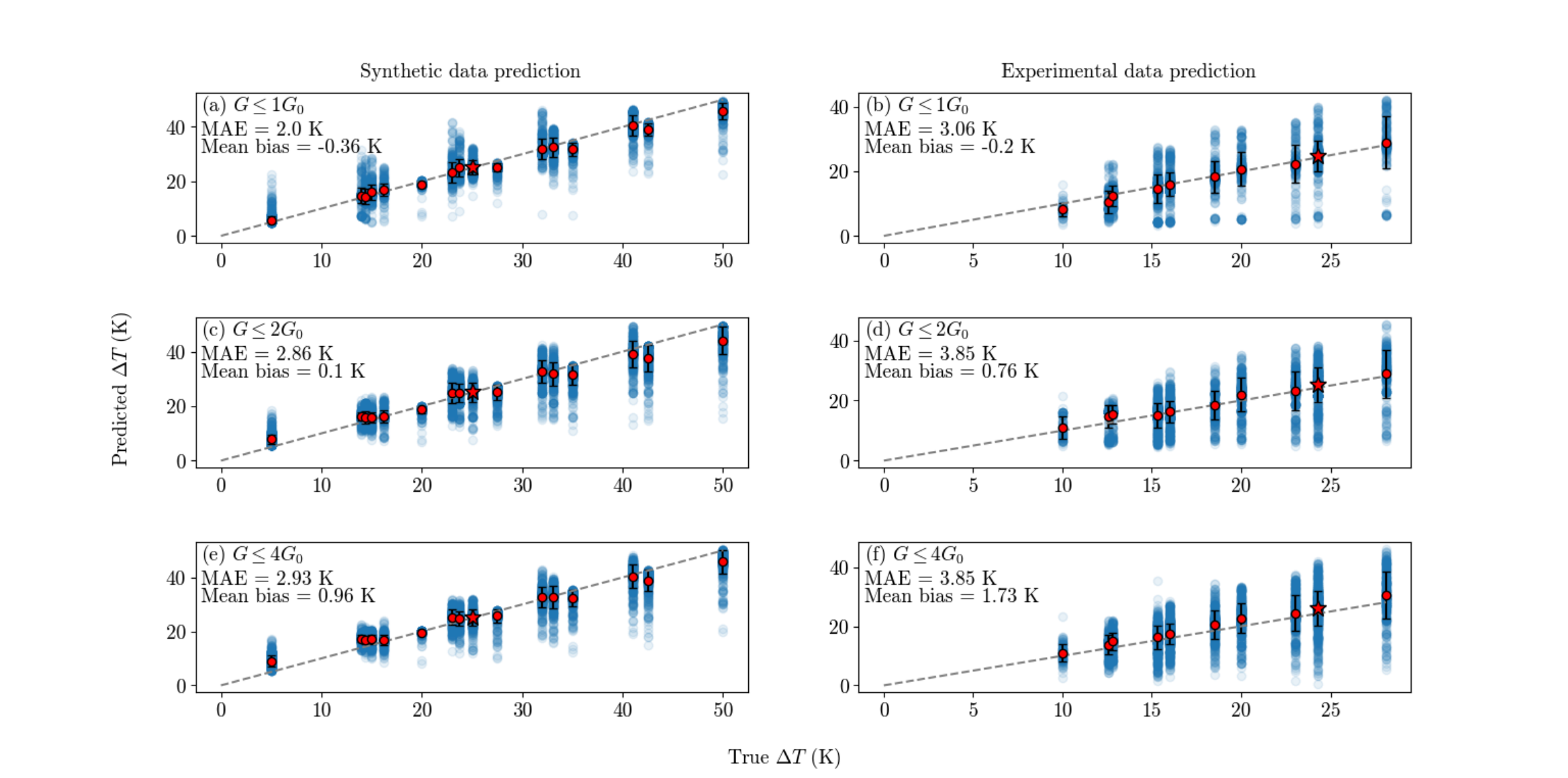}
    \caption{Predictions based on training with the integral formula for delta-T noise, Eq. (\ref{eq:Sor}), 
    after subtracting the equilibrium-like contribution $S_{\bar T}$, which does not depend on $\Delta T$.
    Neural network predicted $\Delta T$ plotted against true $\Delta T$ from datasets generated using Eq. (\ref{eq:Sor}) within the noisy channel opening protocol. Left column panels ((a), (c), (e)) show predictions on the training datasets;  right column panels ((b), (d), (f)) show predictions on the experimental dataset.  Each row presents results with datasets allowing increasingly higher maximum G.
    }
    \label{fig:NCOP-TvT-int}
\end{figure*}

\begin{figure*}[htbp]
    \centering
    \vspace{-3mm}
    \includegraphics[width=1.1\linewidth]{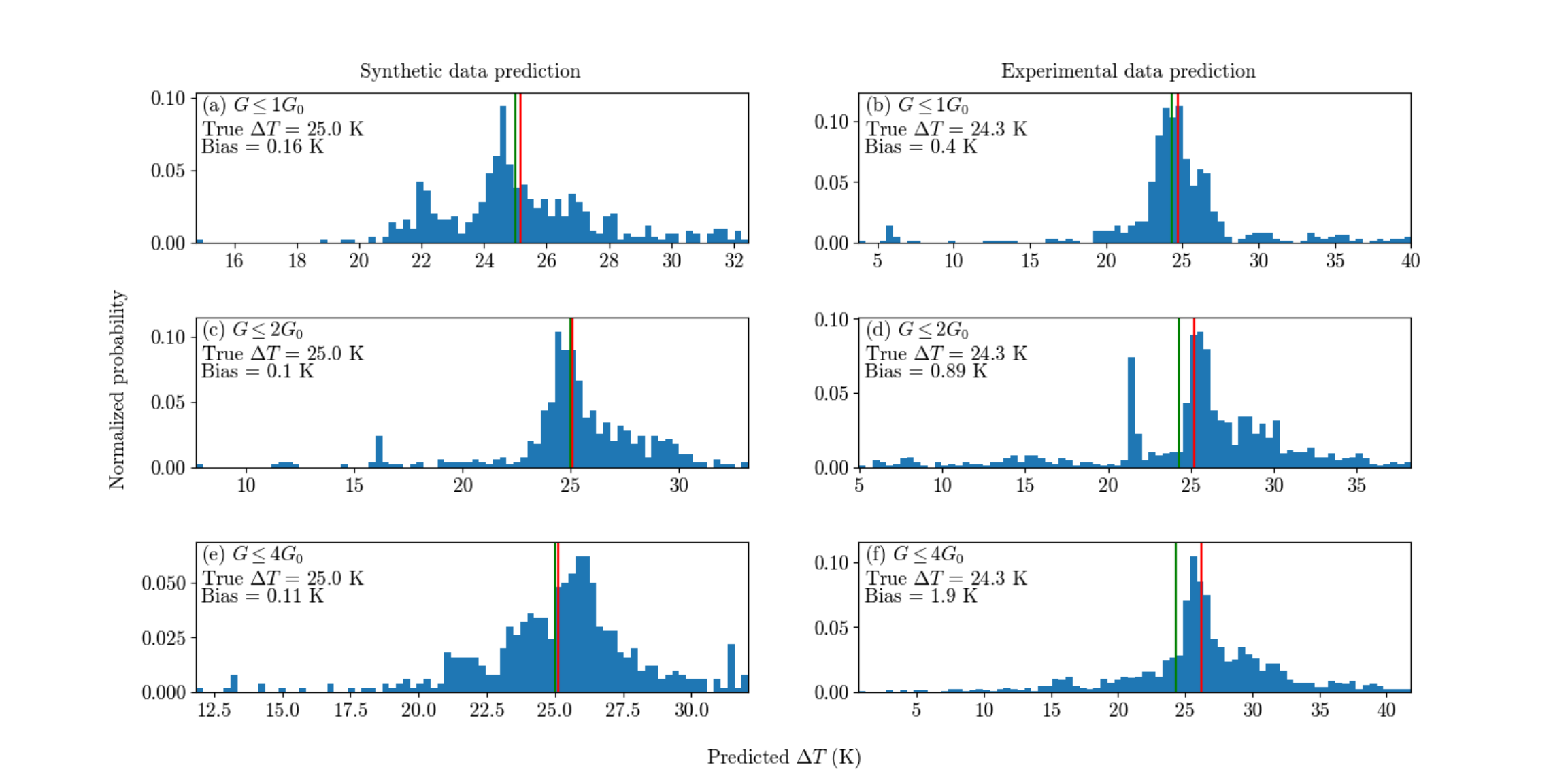}
    \caption{Predictions based on training with the integral formula for delta-T noise. Histograms of particular $\Delta T$ from corresponding panels in Fig. \ref{fig:NCOP-TvT-int}, as indicated there in each panel by red stars. The true $\Delta T$ is shown by the green vertical line; the mean of the histogram is marked by the red line. Left column panels ((a), (c), (e)) share $\bar{T}=17.5$ K and true $\Delta T=25.0$ K. Right column panels ((b), (d), (f)) share $\bar{T}=21.5$ K and true $\Delta T=24.3$ K.}
    \label{fig:NCOP-hist-int}
\end{figure*}

\begin{figure*}[htbp]
    \centering
    \vspace{-3mm}
    \includegraphics[width=1.1\linewidth]{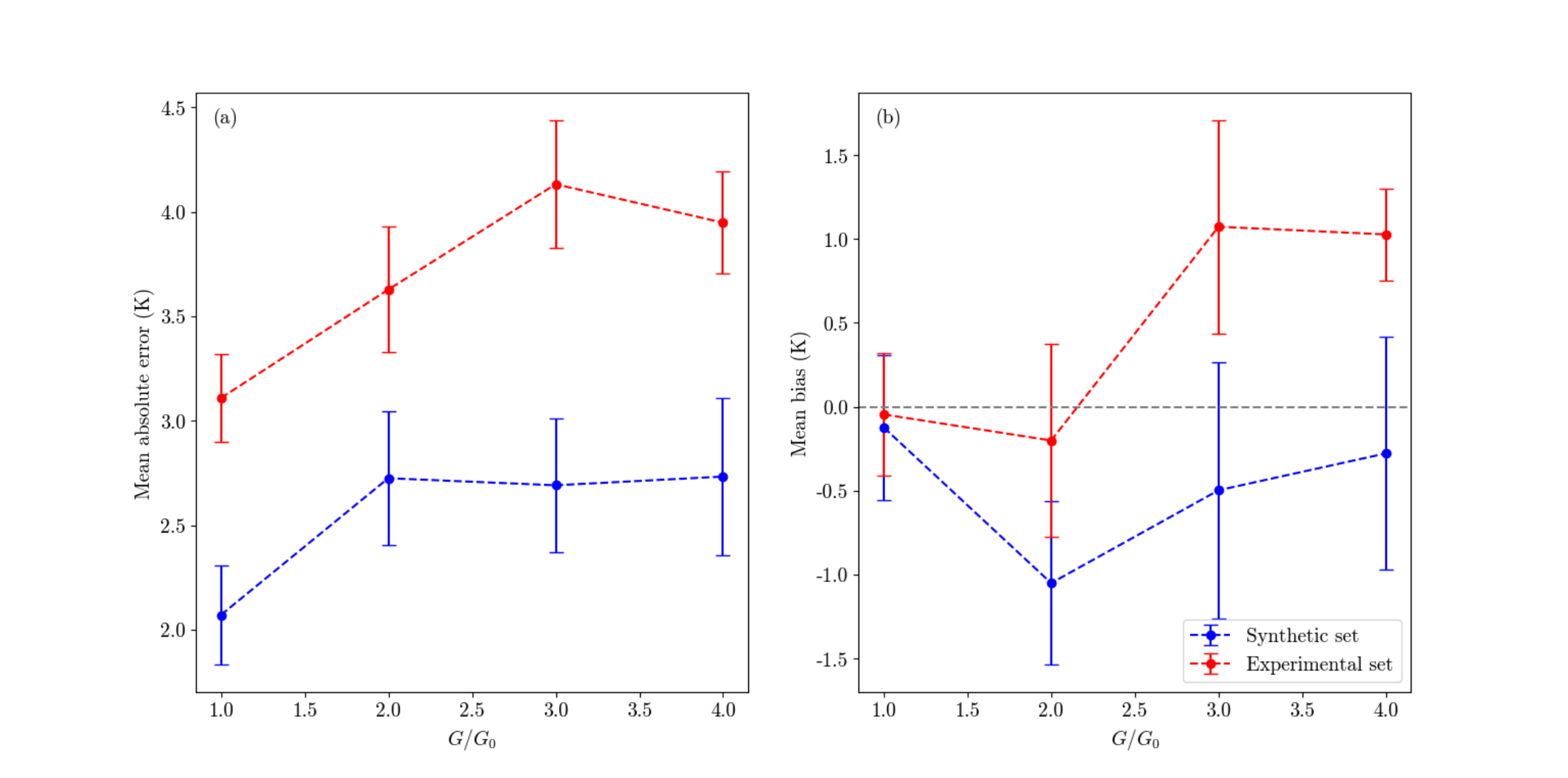}
    \caption{Metrics when training with the integral formula for delta-T noise: (a) mean absolute error and (b) mean bias. Results are shown for $\Delta T$ predictions with varying max $G$ included in datasets. The datasets were generated by the noisy channel opening protocol using the  integral formula (\ref{eq:Sor}). Each data point is averaged over 10 retrained models, and error bars display the standard deviation in the models. Data in blue indicates predictions on synthetic (training) dataset; data in red indicates predictions on experimental dataset.}
    \label{fig:NCOP-G-int}
\end{figure*}

\begin{widetext}
\section{Generating synthetic delta-T noise from the integral noise formula}
\label{sec:App2}

The full counting statistics of charge transport in multi-terminal conductors can be obtained analytically for noninteracting carriers in the coherent limit \cite{Levitov1,Levitov2}. Considering two-terminal junctions ($L$-Left; $R$=right) with a single transmission channel, the charge current is given in the Landauer form,
\bea
I=\frac{2e}{h}\int_{-\infty} ^{\infty} d\epsilon \tau(\epsilon)
[f_L(\epsilon,\mu_L,T_L)-f_R(\epsilon,\mu_R,T_R) ].
\label{eq:IL}
\eea
The current shot noise reads $S_I=S_1+S_2$, where \cite{Rev1}
%
\bea
S_1&=& \frac{4e^2}{h} \int d\epsilon\left\{
f_L(\epsilon,\mu_L,T_L)\left[ 1- f_L(\epsilon,\mu_L,T_L) \right] +
f_R(\epsilon,\mu_R,T_R)\left[ 1- f_R(\epsilon,\mu_R,T_R) \right]
\right\}\tau(\epsilon)^2,
\nonumber\\
S_2&=& \frac{4e^2}{h} \int d\epsilon\left\{
f_R(\epsilon,\mu_R,T_R)\left[ 1- f_L(\epsilon,\mu_L,T_L) \right] +
f_L(\epsilon,\mu_L,T_L)\left[ 1- f_R(\epsilon,\mu_R,T_R) \right]
\right\}\tau(\epsilon)[1- \tau(\epsilon)].
\nonumber\\
\label{eq:Sor}
\eea
%
%
Here, $f_{\nu}(\epsilon,\mu,T)=[ e^{\beta_{\nu}(\epsilon-\mu_{\nu})}+1]^{-1}$ is the Fermi function of the $\nu=L,R$ metal, maintained at temperature $T_{\nu}$ and chemical potential $\mu_{\nu}$, $\tau(\epsilon)$ is the transmission coefficient, which in general varies with the electron energy. In what follows, we assume it is a constant, $\tau$.  The electrical conductance is given by $G=I/V$,
with the applied voltage $V=(\mu_L-\mu_R)/e$. In the limit of constant transmission,  $G=G_0\tau$, with $G_0=2e^2/h$.

Assuming the chemical potentials are identical but that the two metal electrodes are maintained at different temperatures, we expand Eq. (\ref{eq:Sor}) in orders of $\Delta T$. Collecting these contributions, we write the total noise as
\bea S_I&=&S_1+S_2 \nonumber\\
 S_1&=& \frac{4e^2}{h}\mathcal \tau^2
k_B(T_L+T_R),
\nonumber\\
S_2&\approx&\frac{4e^2}{h}\mathcal
\tau(1- \tau) k_B
\Bigg[ \left(T_L+T_R\right) 
+ \frac{(T_L-T_R)^2}{2T}
 \left(
\frac{\pi^2}{9}-\frac{2}{3} \right) \Bigg].
\label{eq:S2b}
\nonumber\\
 \eea
This result leads to Eq. (\ref{eq:S}) in the main text where we decompose the noise to an equilibrium-like contribution, and the excess-nonequilibrium or delta-T noise, $S_I\approx S_{\bar T} + S_{\Delta T}$,
\bea
S_{\bar T}=
\frac{8 e^2}{h} k_B\bar T \tau,
\eea 
and
\bea
S_{\Delta T} = 
\frac{2e^2}{h}\mathcal
\tau(1- \tau) k_B 
 \frac{(T_h-T_c)^2}{\bar T}
 \left(
\frac{\pi^2}{9}-\frac{2}{3} \right). 
\label{eq:SdeltaTA}
\eea
Here and in what follows, we specify the two temperatures by $T_h$ and $T_c$, with their average $\bar T= (T_h+T_c)/2$ and difference $\Delta T=T_h-T_c$.
As for the validity of  Eq. (\ref{eq:SdeltaTA}),
it was shown in Ref. \citenum{DeltaTShot} that the next (quartic) contribution in $\Delta T/\bar T$ was inconsequential relative to the quadratic term, even when using a high temperature difference, $\Delta T = 2\bar T$.
For a multi-channels junctions, $\tau$ is replaced by 
$\sum_i \tau_i$, and $\tau(1 - \tau)$ by
$\sum_i\tau_i(1 - \tau_i)$, with the index $i$ going over the different channels.

In the main text, we utilized the approximate formula (\ref{eq:S}) to generate synthetic data. Here, we  demonstrate that using the more accurate integral formula, Eq. (\ref{eq:Sor}), does not improve the quality of predictions of trained models on experimental datasets, compared to the approximate expression. We assume here constant transmission functions but avoid expanding the Fermi functions near equilibrium. The fact that training with the integral formula does not yield improved performance on experimental data indicates that prediction inaccuracy are not due to 
Eq. (\ref{eq:S}) being an approximation of Eq. (\ref{eq:Sor}) with respect to $\Delta T$.

In Fig. \ref{fig:NCOP-TvT-int}, we train NN on $S_{\Delta T}=S_1+S_2-S_{\bar T}$ with $S_1$ and $S_2$ evaluated from Eq. (\ref{eq:Sor}), using constant transmissions.
We test results on experimental data. Example histograms are displayed in Fig. \ref{fig:NCOP-hist-int}. When comparing these results to predictions shown in Fig. \ref{fig:NCOP-TvT}-\ref{fig:NCOP-hist}, we find that error metrics, as presented in Fig. 
\ref{fig:NCOP-G-int} are similar. 


We highlight that the approximate expression for the delta-T noise, Eq. (\ref{eq:SdeltaTA}) was previously tested only on data up to 1$G_0$ \cite{DeltaTShot}. Here, for the first time, we provide evidence that it  captures behavior beyond that, for higher conductance.  

\end{widetext}




\begin{thebibliography}{4}


\bibitem{MLPC1}
A. L. Ferguson, J. Hachmann, T. F. Miller, and J. Pfaendtner, ``The Journal of Physical Chemistry A/B/C Virtual Special Issue on
Machine Learning in Physical Chemistry," \href{https://doi.org/10.1021/acs.jpcb.0c09206}{J. Phys. Chem. B} {\bf 124}, 9767 (2020).


\bibitem{MLPC2}
A. L. Ferguson and J. Pfaendtner,
``Virtual Special Issue on Machine Learning in Physical Chemistry Volume 2," \href{https://doi.org/10.1021/acs.jpcc.4c03824}{J. Phys. Chem. C} {\bf 128}, 11079 (2024).


\bibitem{RevMO}
P. Gehring, J. M. Thijssen, and H. S. J. van der Zant, ``Single-molecule quantum-transport phenomena in break junctions," \href{https://doi.org/10.1038/s42254-019-0055-1}{Nat. Rev. Phys.} {\bf 1}, 381 (2019). 

\bibitem{Natelsonev}
D. R. Ward, D. A. Corley, J. M. Tour, and D. Natelson, ``Vibrational and electronic heating in nanoscale junctions," \href{https://doi.org/10.1038/nnano.2010.240}{Nat. Nanotechnol.} {\bf 6}, 33 (2011). 

\bibitem{Orenev}
O. Tal, M. Krieger, B. Leerink, and J. M. van Ruitenbeek, ``Electron-Vibration Interaction in Single-Molecule Junctions: From Contact to Tunneling Regimes," \href{https://doi.org/10.1103/PhysRevLett.100.196804}{Phys. Rev. Lett.} {\bf 100}, 196804 (2008).

\bibitem{LathaMM}
M. S. Inkpen, Z.-F. Liu, H. Li, L. M. Campos, J. B. Neaton, and L. Venkataraman, ``Non-chemisorbed gold–sulfur binding prevails in self-assembled monolayers," \href{https://doi.org/10.1038/s41557-019-0216-y}{Nat. Chem.} {\bf 11}, 351 (2019). 

\bibitem{OrenChiral}
A. K. Singh, K. Martin, M. M. Talamo, A. Houssin, N. Vanthuyne, N. Avarvari, and O. Tal, ``Single-molecule junctions map the interplay between electrons and chirality," \href{
https://doi.org/10.48550/arXiv.2408.12258}{arXiv:2408.12258}.


\bibitem{LathaRX}
M. Aziz, C. R. Prindle, W. Lee, B. Zhang, C. Schaack, M. L Steigerwald, F. Zandkarimi, C. Nuckolls, and L. Venkataraman, ``Evaluating the Ability of External Electric Fields to Accelerate Reactions in Solution," \href{https://doi.org/10.1021/acs.jpcb.4c04864}{J. Phys. Chem. B} {\bf 128}, 9553 (2024).




\bibitem{Gemma22}
W. Bro-Jørgensen, J. M. Hamill, R. Bro, and G. C. Solomon, 
``Trusting our machines: validating machine learning models for single-molecule transport experiments," 
\href{https://doi.org/10.1039/D1CS00884F}{Chem. Soc. Rev.} {\bf 51}, 6875 (2022). 

 \bibitem{Taniguchi23}
Y. Komoto, J. Ryu, and M. Taniguchi,
``Machine learning and analytical methods for single-molecule conductance measurements," \href{https://doi.org/10.1039/D3CC01570J}{Chem. Commun.} {\bf 59}, 6796 (2023).

 
\bibitem{Gemma18}
K. P. Lauritzen, A. Magyarkuti, Z. Balogh, A. Halbritter, and G. C. Solomon,
``Classification of conductance traces with recurrent neural networks,"
\href{https://doi.org/10.1063/1.5012514}{J. Chem. Phys.} {\bf 148}, 084111 (2018).


\bibitem{Perrin19}
D. Cabosart, M. El Abbassi, D. Stefani, R. Frisenda, M. Calame, H. S. J. van der Zant, and M. L. Perrin, ``A reference-free clustering method for the analysis of molecular break-junction measurements," \href{https://doi.org/10.1063/1.5089198}{Appl. Phys. Lett.} {\bf 114}, 143102 (2019).  

\bibitem{Vladyka}
A. Vladyka and T. Albrecht, ``Unsupervised classification of single-molecule data with autoencoders and transfer learning," \href{https://iopscience.iop.org/article/10.1088/2632-2153/aba6f2}{Mach. Learn.: Sci. Technol.} {\bf 1}, 035013 (2020).  

\bibitem{Hong20}
F. Huang, R. Li, G. Wang, J. Zheng, Y. Tang, J. Liu, Y. Yang, Y. Yao, J. Shi, and W. Hong, ``Automatic classification of single-molecule charge transport data with an unsupervised machine-learning algorithm," \href{https://doi.org/10.1039/C9CP04496E}{Phys. Chem. Chem. Phys.} {\bf 22}, 1674 (2020).  

\bibitem{Hong21}
L. Lin, C. Tang, G. Dong, Z. Chen, Z. Pan, J. Liu, Y. Yang, J. Shi, R. Ji, and W. Hong, ``Spectral Clustering to Analyze the Hidden Events in Single-Molecule Break Junctions," \href{https://doi.org/10.1021/acs.jpcc.0c11473}{J. Phys. Chem. C} {\bf 125}, 3623 (2021).  

\bibitem{Zant19}
M. El Abbassi, P. Zwick, A. Rates, D. Stefani, A. Prescimone, M. Mayor, H. S. J. van der Zant, and D. Duli\'{c},
``Unravelling the conductance path through single-porphyrin junctions," \href{https://doi.org/10.1039/C9SC02497B}{Chem. Sci.} {\bf 10}, 8299 (2019).


\bibitem{Zant21}
M. El Abbassi, J. Overbeck, O. Braun, M. Calame, H. S. J. van der Zant, and M. L. Perrin, ``Benchmark and application of unsupervised classification approaches for univariate data," \href{https://doi.org/10.1038/s42005-021-00549-9}{Commun. Phys.}  {\bf 4}, 50 (2021).  


\bibitem{Latha20}
T. Fu, Y. Zang, Q. Zou, C. Nuckolls, and L. Venkataraman, ``Using Deep Learning to Identify Molecular Junction Characteristics," \href{https://doi.org/10.1021/acs.nanolett.0c00198}{Nano Lett.} {\bf 20}, 3320 (2020). 


\bibitem{Gemma24}
W. Bro-Jørgensen, J. M. Hamill, G. Mezei, B. Lawson, U. Rashid, A. Halbritter, M. Kamenetska, V. Kaliginedi, G. C. Solomon,
``Making the Most of Nothing: One-Class Classification for Single-Molecule Transport Studies," 
\href{https://doi.org/10.1021/acsnanoscienceau.4c00015}{ACS Nanosci. Au} {\bf 4}, 250 (2024).

\bibitem{Anatole14}
A. Lopez-Bezanilla and O. A. von Lilienfeld,
``Modeling electronic quantum transport with machine learning,"
\href{https://doi.org/10.1103/PhysRevB.89.235411}{Phys. Rev. B} {\bf 89}, 235411 (2014).


\bibitem{Li20}
K. Li, J. Lu, and F. Zhai,
``Neural networks for modeling electron transport properties of mesoscopic systems,"
\href{https://doi.org/10.1103/PhysRevB.102.064205}{Phys. Rev. B} {\bf 102}, 064205 (2020).

\bibitem{RomanDNA}
R. Korol and D. Segal,
``Machine Learning Prediction of DNA Charge Transport,"
\href{https://doi.org/10.1021/acs.jpcb.8b12557}{J. Phys. Chem. B} {\bf 123}, 2801 (2019).

\bibitem{MaitiDNA}
A. Aggarwal, V. Vinayak, S. Bag, C. Bhattacharyya, U. V. Waghmare, and P. K. Maiti,
``Predicting the DNA Conductance Using a Deep Feedforward Neural Network Model," \href{https://doi.org/10.1021/acs.jcim.0c01072}{J. Chem. Inf. Model.} {\bf 61}, 106 (2021). 


\bibitem{FP21}
M. Bürkle, U. Perera, F. Gimbert, H. Nakamura, M. Kawata, and Y. Asai, ``Deep-Learning Approach to First-Principles Transport Simulations," \href{https://doi.org/10.1103/PhysRevLett.126.177701}{Phys. Rev. Lett.} {\bf 126}, 177701 (2021).


\bibitem{MDML}
R. Topolnicki, R. Kucharczyk, and W. Kaminski,
``Combining Multiscale MD Simulations and Machine Learning
Methods to Study Electronic Transport in Molecular Junctions at
Finite Temperatures,"
\href{https://doi.org/10.1021/acs.jpcc.1c03210}{J. Phys. Chem. C} {\bf 125}, 19961 (2021).


\bibitem{Franco23}
M. Deffner, M. P. Weise, H. Zhang, M. M\"{u}cke, J. Proppe, I. Franco, and C. Herrmann,
``Learning Conductance: Gaussian Process Regression for Molecular Electronics," \href{https://doi.org/10.1021/acs.jctc.2c00648}{J. Chem. Theory Comput.} {\bf 19}, 992 (2023).

\bibitem{Ilia1}
I. Khait, J. Carrasquilla, and D. Segal,
``Optimal control of quantum thermal machines using machine learning,"
\href{https://doi.org/10.1103/PhysRevResearch.4.L012029}{Phys. Rev. Res.} {\bf 4}, L012029 (2022).

\bibitem{Erdman1}
P. A Erdman and F. Noé, 
``Model-free optimization of power/efficiency tradeoffs in quantum thermal machines using reinforcement learning," 
\href{https://doi.org/10.1093/pnasnexus/pgad248}{PNAS Nexus} {\bf 2}, pgad248 (2023).

\bibitem{Erdman2}
P. A. Erdman and F. Noé,
``Identifying optimal cycles in quantum thermal machines with reinforcement-learning,"
\href{https://doi.org/10.1038/s41534-021-00512-0}{NPJ Quantum Inf.} {\bf 8}, 1 (2022).

\bibitem{Erdman3}
F. Mazzoncini, V. Cavina, G. M. Andolina, P. A. Erdman, and V. Giovannetti,
``Optimal control methods for quantum batteries,"
\href{https://doi.org/10.1103/PhysRevA.107.032218}{Phys. Rev. A} {\bf 107}, 032218 (2023).

\bibitem{Ilia2}
S. Saryal, M. Gerry, I. Khait, D. Segal, and B. K. Agarwalla, 
``Universal bounds on fluctuations in continuous thermal machines,"
\href{https://doi.org/10.1103/PhysRevLett.127.190603}{Phys. Rev. Lett.} {\bf 127}, 190603 (2021).


\bibitem{DeltaTShot}
O. Shein Lumbroso, L. Simine, A. Nitzan, D. Segal, and O. Tal, ``Electronic noise due to temperature difference in atomic-scale junctions,"
\href{https://doi.org/10.1038/s41586-018-0592-2}{Nature} {\bf 562}, 240 (2018).


\bibitem{TunnelJ}
S. Larocque, E. Pinsolle, C. Lupien, and B. Reulet,
``Shot Noise of a Temperature-Biased Tunnel Junction,"
\href{https://doi.org/10.1103/PhysRevLett.125.106801}{Phys. Rev. Lett.} {\bf 125}, 106801 (2020).


\bibitem{Levitov1}
H. Lee, L. S. Levitov, and A. Y. Yakovets, ``Universal statistics of transport in disordered conductors," \href{https://doi.org/10.1103/PhysRevB.51.4079}{Phys. Rev. B} {\bf 51}, 4079 (1995). 

\bibitem{Levitov2}
L. S. Levitov, H.-W. Lee, and G. B. Lesovik, ``Electron counting statistics and coherent states of electric current," \href{https://doi.org/10.1063/1.531672}{J. Math. Phys.} {\bf 37}, 4845 (1996).


\bibitem{Rev1}
Y. M. Blanter and M. B\"uttiker, ``Shot noise in mesoscopic conductors," 
\href{https://doi.org/10.1016/S0370-1573(99)00123-4}{Phys. Rep.} {\bf 336}, 1 (2000).


\bibitem{AnqiR}
A. Mu and D. Segal, ``Anomalous electronic shot noise in resonant tunneling junctions," \href{https://doi.org/10.48550/arXiv.1902.06312}{arXiv:1902.06312}.

\bibitem{qdot}
A. Popoff, J. Rech, T. Jonckheere, L. Raymond, B. Grémaud, S. Malherbe, and T. Martin,
``Scattering theory of non-equilibrium noise and delta T current fluctuations through a quantum dot,"
\href{https://iopscience.iop.org/article/10.1088/1361-648X/ac5200/meta}{J. Phys.: Condens. Matter} {\bf 34}, 185301 (2022).

\bibitem{Janine1}
J. Eriksson, M. Acciai, L. Tesser, and J. Splettstoesser,
``General Bounds on Electronic Shot Noise in the Absence of Currents,"
\href{https://doi.org/10.1103/PhysRevLett.127.136801}{Phys. Rev. Lett.} {\bf 127}, 136801 (2021).


\bibitem{light}
M. Hübler and W. Belzig,
``Light emission in delta-$T$-driven mesoscopic conductors,"
\href{https://doi.org/10.1103/PhysRevB.107.155405}{Phys. Rev. B} {\bf 107}, 155405 (2023).

\bibitem{Janine2}
L. Tesser, M. Acciai, C. Spånslätt, J. Monsel, and J. Splettstoesser,
``Charge, spin, and heat shot noises in the absence of average currents: Conditions on bounds at zero and finite frequencies,"
\href{https://doi.org/10.1103/PhysRevB.107.075409}{Phys. Rev. B} {\bf 107}, 075409 (2023).

\bibitem{SuperC}
Leonardo Pierattelli, Fabio Taddei, Alessandro Braggio,
``$\Delta T$-noise in Multiterminal Hybrid Systems,"
\href{https://doi.org/10.48550/arXiv.2411.12572}{arXiv:2411.12572}.


\bibitem{Hall}
J. Rech, T. Jonckheere, B. Grémaud, and T. Martin,
``Negative Delta-$T$ Noise in the Fractional Quantum Hall Effect,"
\href{https://doi.org/10.1103/PhysRevLett.125.086801}{Phys. Rev. Lett.} {\bf 125}, 086801 (2020).

\bibitem{Kondo}
M. Hasegawa and K. Saito,
``Delta-$T$ noise in the Kondo regime,"
\href{https://doi.org/10.1103/PhysRevB.103.045409}{Phys. Rev. B} {\bf 103}, 045409 (2021).

\bibitem{FHall}
G. Rebora, J. Rech, D. Ferraro, T. Jonckheere, T. Martin, and M. Sassetti,
``Delta-$T$ noise for fractional quantum Hall states at different filling factor,"
\href{https://doi.org/10.1103/PhysRevResearch.4.043191}{Phys. Rev. Research} {\bf 4}, 043191 (2022).

\bibitem{color}
K. Iyer, J. Rech, T. Jonckheere, L. Raymond, B. Grémaud, and T. Martin,
``Colored delta-$T$ noise in fractional quantum Hall liquids,"
\href{https://doi.org/10.1103/PhysRevB.108.245427}{Phys. Rev. B} {\bf 108}, 245427 (2023).



\bibitem{Rev2}
S. Yuan, T. Gao, W. Cao, Z. Pan, J. Liu, J. Shi, W. Hong, ``The Characterization of Electronic Noise in the Charge Transport through Single-Molecule Junctions," 
\href{https://doi.org/10.1002/smtd.202001064}{Small Methods} {\bf 5}, 2001064 (2021). 

\bibitem{Exp}
S. Tewari, C. Sabater, M. Kumar, S. Stahl, B. Crama, and J. M. van Ruitenbeek,
``Fast and accurate shot noise measurements on atomic-size junctions in the MHz regime,"
\href{https://doi.org/10.1063/1.5003391}{Rev. Sci. Instrum.} {\bf 88} 093903 (2017).

\bibitem{Vardimon2013}
R. Vardimon, M. Klionsky, and O. Tal,
``Experimental determination of conduction channels in atomic-scale conductors based on shot noise measurements,"
\href{https://doi.org/10.1103/PhysRevB.88.161404}{Phys. Rev. B} {\bf 88}, 161404 (2013).


\bibitem{Berndt19}
M. Mohr, T. Jasper-Toennies, A. Weismann, T. Frederiksen, A. Garcia-Lekue, S. Ulrich, R. Herges, and R. Berndt, 
``Conductance channels of a platform molecule on Au(111) probed with shot noise,"
\href{https://doi.org/10.1103/PhysRevB.99.245417}{Phys. Rev. B} {\bf 99}, 245417 (2019).

\bibitem{Berndt20}
M. Mohr, M. Gruber, A. Weismann, D. Jacob, P. Abufager, N. Lorente, and R. Berndt,
``Spin dependent transmission of nickelocene-Cu contacts probed with shot noise,"
\href{https://doi.org/10.1103/PhysRevB.101.075414}{Phys. Rev. B} {\bf 101}, 075414 (2020).

\bibitem{Sheer16}
M. A. Karimi, S. G. Bahoosh, M. Herz, R. Hayakawa, F. Pauly, and E. Scheer,
``Shot Noise of 1,4-Benzenedithiol Single-Molecule Junctions,"
\href{https://doi.org/10.1021/acs.nanolett.5b04848}{Nano Lett.} {\bf 16}, 1803 (2016).

\bibitem{Sheer21}
M. W. Prestel, M. Strohmeier, W. Belzig, and E. Scheer,
``Revealing channel polarization of atomic contacts of ferromagnets and strong paramagnets by shot-noise measurements,"
\href{https://doi.org/10.1103/PhysRevB.104.115434}{Phys. Rev. B} {\bf 104}, 115434 (2021).


\bibitem{Natelson1}
R. Chen, P. J. Wheeler, M. Di Ventra, and D. Natelson,
``Enhanced noise at high bias in atomic-scale Au break junctions,"
\href{https://doi.org/10.1038/srep04221}{Sci. Rep.} {\bf 4}, 4221 (2014).


\bibitem{Natelson2}
R. Chen and D. Natelson,
``Evolution of shot noise in suspended lithographic gold break junctions with bias and temperature,"
\href{https://iopscience.iop.org/article/10.1088/0957-4484/27/24/245201}{Nanotechnology} {\bf 27}, 245201 (2016).

\bibitem{Latha15}
O. Adak, E. Rosenthal, J. Meisner, E. F. Andrade, A. N. Pasupathy, C. Nuckolls, M. S. Hybertsen, and L. Venkataraman, ``Flicker noise as a probe of electronic interaction at metal-single molecule interfaces," 
\href{https://doi.org/10.1021/acs.nanolett.5b01270}{Nano Lett.} {\bf 15}, 4143 (2015).

\bibitem{Anqi}
A. Mu, O. Shein-Lumbroso, O. Tal, and D. Segal,
``Origin of the Anomalous Electronic Shot Noise in Atomic-Scale Junctions,"
\href{https://doi.org/10.1021/acs.jpcc.9b06766}{J. Phys. Chem. C} {\bf 123}, 23853 (2019).

\bibitem{Nitzanev}
M. Galperin, A. Nitzan, and M. A. Ratner,
``Inelastic tunneling effects on noise properties of molecular junctions,"
\href{https://doi.org/10.1103/PhysRevB.74.075326}{Phys. Rev. B} {\bf 74}, 075326 (2006).

\bibitem{lightN}
K. Kaasbjerg and A. Nitzan,
``Theory of light emission from quantum noise in plasmonic contacts: Above-threshold emission from higher-order electron-plasmon scattering,"
\href{https://doi.org/10.1103/PhysRevLett.114.126803}{Phys. Rev. Lett.} {\bf 114}, 126803 (2015).

\bibitem{Bijayev}
B. K. Agarwalla, J. H. Jiang, and D. Segal,
``Full counting statistics of vibrationally assisted electronic conduction: Transport and fluctuations of thermoelectric efficiency,"
\href{https://doi.org/10.1103/PhysRevB.92.245418}{Phys. Rev. B} {\bf 92}, 245418 (2015).
%




\bibitem{Latha24}
A. L. Paoletta and L. Venkataraman, 
``Determining Transmission Characteristics from Shot-Noise-Driven Electroluminescence in Single-Molecule Junctions,"
\href{https://doi.org/10.1021/acs.nanolett.3c04207}{Nano Lett.} {\bf 24}, 1931 (2024).

\bibitem{ruitenbeek1999}
H. E. van den Brom and J. M. van Ruitenbeek,
``Quantum Suppression of Shot Noise in Atom-Size Metallic Contacts,"
\href{https://doi.org/10.1103/PhysRevLett.82.1526}{Phys. Rev. Lett.} {\bf 82}, 1526(4) (1999).


\bibitem{Stafford}
J. Bürki and C. A. Stafford,
``Comment on ``Quantum Suppression of Shot Noise in Atom-Size Metallic Contacts,"
\href{https://doi.org/10.1103/PhysRevLett.83.3342}{Phys. Rev. Lett.} 
{\bf 83}, 3342 (1999).

\bibitem{pauly2011}
F. Pauly, J. K. Viljas, M. Burkle, M. Dreher, P. Nielaba, and J. C. Cuevas,
``Molecular dynamics study of the thermopower of Ag, Au, and Pt nanocontacts,"
\href{https://doi.org/10.1103/PhysRevB.84.195420}{Phys. Rev. B} {\bf 84}, 195420 (2011).

\bibitem{Cuevas2017}
J. C. Kl\"{o}ckner, M. Matt, P. Nielaba, F. Pauly, and J. C. Cuevas,
``Thermal conductance of metallic atomic-size contacts: Phonon transport and Wiedemann-Franz law,"
\href{https://doi.org/10.1103/PhysRevB.96.205405}{Phys. Rev. B} {\bf 96}, 205405 (2017).

\bibitem{Keras}
F. Chollet, ``Keras", \href{https://github.com/fchollet/keras}{GitHub}, (2015).

\bibitem{flicker2}
O. Shein-Lumbroso, M. Gerry, A. Shastry, A. Vilan, D. Segal, and O. Tal, ``Delta-T flicker noise demonstrated with molecular junctions," 
\href{https://doi.org/10.1021/acs.nanolett.3c04445}{Nano Lett.} {\bf 24}, 1981 (2024).

\bibitem{Adam}
D. P. Kingma and J. Ba,
``Adam: A Method for Stochastic Optimization,"
\href{https://doi.org/10.48550/arXiv.1412.6980}{arXiv:1412.6980}.



























\end{thebibliography}
\end{document}